\documentclass[aps,prd,onecolumn,superscriptaddress,groupedaddress,nofootinbib]{revtex4}
\usepackage[utf8]{inputenc}
\usepackage[T1]{fontenc}
\usepackage{graphicx}
\usepackage{epsfig}
\usepackage{amsmath,amssymb,amsthm}
\usepackage{caption}
\usepackage{subcaption}
\usepackage[normalem]{ulem}
\usepackage{xcolor}
\usepackage{tikz}
\usepackage[compat=1.1.0]{tikz-feynman}
\usepackage{hyperref}
\usepackage{multirow}

\newcommand\Rsout{\bgroup\markoverwith{\textcolor{red}{\rule[0.5ex]{2pt}{0.4pt}}}\ULon}
\newcommand\Bsout{\bgroup\markoverwith{\textcolor{blue}{\rule[0.5ex]{2pt}{0.4pt}}}\ULon}

\newcommand{\Eq}[1]{Eq.(#1)}
\newcommand{\mcm}{\ensuremath{m_s}}

\newcommand{\DqC}[2]{\ensuremath{\mathcal{D}_4\hspace{-0.1cm}\begin{pmatrix} #1 & #1 & 0 \\ c & 1 & #2\end{pmatrix}}}
\newcommand{\DqCfull}[6]{\ensuremath{\mathcal{D}_4\hspace{-0.1cm}\begin{pmatrix} #1 & #2 & #3 \\ #4 & #5 & #6\end{pmatrix}}}
\newcommand{\DqTC}[3]{\ensuremath{\Tilde{\mathcal{D}}_4\hspace{-0.1cm}\begin{pmatrix} #1 & \Big\rvert & #2 & #2 & 0\\ \varepsilon & \Big\rvert & c & 1 & #3\end{pmatrix}}}

\newcommand{\DqTCC}[2]{\ensuremath{\Tilde{\mathcal{D}}_4\hspace{-0.1cm}\begin{pmatrix} #1 & \Big\rvert & #2 & #2 & 0\\ \varepsilon & \Big\rvert & c & c & \varepsilon\end{pmatrix}}}

\newcommand{\DcT}[3]{\ensuremath{\Tilde{\mathcal{D}}_5\hspace{-0.1cm}\begin{pmatrix} #1 &\Big\rvert & #2\\ \varepsilon & \Big\rvert  & #3\end{pmatrix}}}
\newcommand{\e}{\ensuremath{\varepsilon}}
\newcommand{\Ll}[1]{\ensuremath{\ln\hspace{-0.05cm}\frac{\Lambda}{m_{#1}}}}
\newcommand{\Llc}[1]{\ensuremath{\ln^2\hspace{-0.05cm}\frac{\Lambda}{m_{#1}}}}
\newcommand{\be}{\begin{equation}}
\newcommand{\ee}{\end{equation}}
\newcommand{\msbar}{{\overline{\mbox{\rm MS}}}}
\newcommand{\ba}{\begin{eqnarray}}
\newcommand{\ea}{\end{eqnarray}}
\newcommand{\nn}{\nonumber \\}
\newcommand{\pmq}{a}
\newcommand{\fr}[2]{{\frac{#1}{#2}\,}}
\renewcommand{\(}{\left(}
\renewcommand{\)}{\right)}
\renewcommand{\[}{\left[}
\renewcommand{\]}{\right]}

\newcommand{\mh}{\ensuremath{\hat{m}}}
\newcommand{\uh}{\ensuremath{\hat{u}}}
\newcommand{\nui}{\ensuremath{\nu_{ij}}}

\newcommand{\as}{\ensuremath{\alpha_S}}

\begin{document}
\title{Cold Quark Matter: Renormalization Group improvement \\ at next-to-next-to leading order}
\author{Loïc Fernandez}
\affiliation{Helsinki Institute of Physics, P.O. Box 64, FI-00014 University of Helsinki, Finland}
\author{Jean-Loïc Kneur} 
\affiliation{Laboratoire Charles Coulomb (L2C), UMR 5221 Universit\'e de Montpellier, France}

\begin{abstract}
We extend previous next-to-next-to leading order (NNLO) calculations of the QCD pressure at zero temperature and non-zero baryonic densities
using the renormalization group optimized perturbation theory (RGOPT), which entails
an all-order RG-invariant resummation. First, we consider the approximation of three massless quark flavors, and then we add the running 
strange quark mass dependence.
The resulting pressure displays a sizeably reduced sensitivity to variations of the arbitrary renormalization scale as compared to the state-of-the-art NNLO results. 
This confirms previous NLO investigations that the RGOPT resummation scheme provides improved convergence properties 
and reduced renormalization scale uncertainties, thus being a promising
prescription to improve perturbative QCD at high and mid range baryonic densities.
\end{abstract} 
\maketitle

\section{Introduction}\label{Introduction}
Quantum Chromodynamics (QCD) displays nonperturbative features, even at weak coupling values, 
when either finite temperature ($T$)
and/or baryonic density are taken into consideration. This complication is due to the appearance of long-range correlations in the medium for gluonic fields (see \cite{Trev} for reviews), making any analytical approach to the determination of the partition function and associated thermodynamical quantities challenging.
For vanishing baryonic densities and high temperatures, the state-of-the-art thermodynamical calculations\cite{lattice} are held by lattice QCD (LQCD) simulations, however, owing to the
sign problem in the fermionic sector\cite{sign}, presently LQCD cannot explore the mid and high range of baryonic chemical potential 
($\mu_B$) values. At low baryonic densities, chiral perturbation theory gives a quite accurate description\cite{Drischler:2017wtt,chptdense},
whereas perturbative QCD (pQCD) is reliable only for relatively high chemical potential values, 
leaving the mid range, particularly relevant\cite{NSrev} to the description of Neutron Stars (NS), still uncertain. 
Recently developed approaches were used in order to reconnect the lower and higher baryonic density ranges, in particular with 
model-independent inferences for the equations of state (EoS) of compact stars (see e.g. \cite{vuorNS,KurkelaNS,Annala:2023cwx,Komoltsev:2023zor}). 
Moreover, QCD is expected to undergo a phase transition\cite{critrev} for some intermediate value of $\mu_B$, thus further complexifying an accurate EoS description. While Heavy Ion collision experiments at RHIC and LHC 
have shown evidence for a Quark-Gluon-Plasma (QGP)\cite{CBMrev}, 
present and future experiments\cite{CBMrev} 
in particular FAIR\cite{FAIR}, aim to systematically scan the finite density region of the phase diagram looking for a postulated critical endpoint separating the first order phase transition from the smooth crossover\cite{lattice} region.

Thanks to the asymptotic freedom property of QCD, one could hope at first that a perturbative expansion could be reliable. However, it is well known that at least two different scales naturally emerge in thermal field theory: the hard scale $T$  (resp. $\mu_B$) and the soft scale $g\,T$  (resp. $g\,\mu_B$)
at finite T (respectively finite $\mu$ and $T\to 0$)\footnote{At finite $T$ there is in addition  
the ultrasoft scale $g^2 T$, associated with the Linde\cite{Linde} nonperturbative problem, but the latter is not present 
 at $T=0$ and $\mu\neq 0$.}.  When bosonic 
 degrees of freedom are present, the soft scale is responsible for the apparition of long range interactions in the medium, which cannot be consistently reproduced by a naive weak-coupling expansion. This property is usually seen in a massless gauge field which develops an infinity of nonphysical infrared divergences (IR) in Feynman diagrams. 
 These IR divergences have to be appropriately resummed to give reliable predictions, and lead to non-analytical coupling dependence in the pressure and other resulting thermodynamical quantities. At finite $\mu_B$ but zero temperature, the perturbative expansion suffers relatively less instability than its counterpart at finite $T$ and $\mu_B=0$. The pioneering calculation by Freedman and McLerran\cite{soft1} of the next-to-next-to-leading order (NNLO) pressure at vanishing temperature and finite chemical potential (in the massless quark approximation), displayed the emergence of an $\as^2 \ln \as$ dependence  from the plasmon (ring) resummation of soft divergences.
More recent efforts lead to a full calculation of the pQCD pressure at NNLO for finite 
quark masses\cite{Fraga:2004gz,LaineSchroder,Kurkela:2009gj,Fraga_2014} including thermal effects \cite{PhysRevD.74.045016}\cite{PhysRevLett.117.042501}.
A successful alternative resummation method is to systematically expand 
from the start about a quasiparticle mass, which naturally circumvent 
infrared divergences, like is done in screened perturbation theory (SPT)\cite{spt1,spt3}, or concerning QCD,
the Hard Thermal Loop perturbation theory (HTLpt)\cite{HTLpt}, based on the HTL effective theory\cite{HTLbasic}. Expanding about a quasiparticle mass is reminiscent of resummation approaches also 
used at zero temperature and density, like the traditional Hartree approximation and its 
variational generalizations. Essentially, it implies that the original Lagrangian is modified by a Gaussian mass term, to be treated as an interaction. Already at NLO, one usually goes beyond the simple Hartree
approximation since the variational mass is dressed by incorporating different 
resummed topologies (exchange terms, vertex corrections, etc) order by order. This results in a sequence of ``variationally improved'' approximations at successive orders.

From the renormalization group (RG) properties we know that the QCD free energy should be 
an RG-invariant quantity in the complete theory, but practical calculations are limited to
the first few orders of perturbative
expansion,  spoiling this property, and a residual renormalization scale dependence is unavoidable. The latter is conventionally often taken
as a measure of the theoretical uncertainties due to unknown higher
perturbative orders.
Accordingly, upon including higher order terms from the weak-coupling expansion, one expect this 
residual scale dependence to decrease. However, it has been observed in the context of HTLpt 
\cite{HTLpt,HTLpt2g,HTLptDense2L,HTLptDense3L}, notably 
at finite temperatures, that the residual scale sensitivity remains important, and even increases when successive terms in the 
weak-coupling expansion are considered. In contrast, for $T=0$ (cold and dense) quark matter, the residual scale dependence uncertainties of the state-of-the-art results for thermodynamics quantities (i.e., NNLO for massive quarks \cite{Kurkela:2009gj} and $\rm N^3LO$ in the massless quark approximation\cite{Gorda:2018gpy,Gorda:2021znl,Gorda:2021kme,Gorda:2023mkk})
are milder than for their counterpart quantities at $T \ne 0$. Moreover, the 
scale dependence decreases upon including more and more higher order 
contributions\cite{Gorda:2023mkk}.
Those $T=0,\mu \ne 0$ results 
are actually perturbatively RG invariant, namely the residual scale dependence 
formally enters explicitly only at higher ($\rm N^3LO$) order.
Nevertheless, those results still exhibit a rather sizable scale dependence, specially 
at intermediate and rather low $\mu_B$ values and for the
massive quark case, hinting that
there is room for further improvement from RG resummation properties, as indeed exemplified
recently\cite{Fernandez:2021jfr}. 

In the present work, we consider the NNLO pressure for three flavors of quarks at vanishing temperature and high baryonic density, using an alternative resummation method: the Renormalization Group Optimized perturbation theory (RGOPT)\cite{RGOPTals}.
This approach was initially a RG-improved variant of Optimized Perturbation Theory (OPT)\footnote{There is a vast literature on OPT and
its variants, see e.g. \cite{opt_pms,odm,opt_LDE,optT}.}, at vanishing temperature and density. In contrast with standard perturbation theory, RGOPT  can
provide nontrivial results for the order parameters
of chiral symmetry breaking, in particular for the quark condensate\cite{KNcond,KNcond2}, 
in excellent agreement with lattice simulation results.
When applied at finite temperatures, RGOPT
can be considered a RG-improved variant of SPT and HTLpt, which crucially involves a thermal 
(medium)-dressed screening mass. The RGOPT has been investigated at finite temperature first for the $\lambda\phi^{4}$ theory \cite{rgopt_phi4}
up to NNLO \cite{rgopt_phi4_NNLO}, and at NLO for the QCD pressure\cite{rgopt_hot}, where in both cases it was shown to drastically reduce
the residual scale dependence, as compared to standard perturbation theory  and HTLpt. Moreover, for hot QCD 
this NLO approximation is quite close to lattice simulation results\cite{lattice} (at least down to
temperatures moderately above $T_c$).
Concerning cold quark matter, RGOPT was investigated at NLO\cite{rgopt_cold}, 
producing again a reduced residual scale dependence (although with more moderate effect than
for the $T\ne 0$ case). Therefore, in this work our main purpose is to extend the RGOPT approach to cold quark matter
at NNLO. As compared to NLO\cite{rgopt_cold}, there are more challenging issues to cope with
when incorporating our RG approach, due to the quite involved quark mass dependence, as obtained
perturbatively from the state-of-the-art standard weak-coupling expansion.

The paper is organized as follows: In Section II, 
we review the standard NNLO
cold and dense pressure relevant for two massless and one massive (strange) quark flavor (referred to as $N_f=2+1^*$), originally obtained in \cite{Kurkela:2009gj}. 
Due to needed generalizations for the RGOPT construction as detailed below, 
we had to reconsider from basics the latter NNLO calculation, and we fully reproduce all the analytical results from \cite{Kurkela:2009gj}.
We obtain, however, some numerical differences in the quark mass dependence of 
one fitting function introduced in \cite{Kurkela:2009gj}. 
As we will illustrate, the differences produce hardly visible effects in the resulting standard NNLO pressure for actual values of the physical strange quark mass,
while it has potentially more sizable impact for larger quark masses, in particular
for the ``medium-dressed'' quark masses entering our subsequent RGOPT construction. In Section III, we generalize the NNLO pressure expression to $N_f=3^*$ (symmetric, i.e. degenerate) massive quarks, a needed ingredient of the RGOPT construction. 
The perturbative modifications of standard weak-coupling expansion results as implied by the RGOPT approach are reviewed and worked out to evaluate the quark contribution to the NNLO RGOPT pressure at vanishing temperatures and finite baryonic densities. In section IV we thus present our NNLO results for this symmetric case, compared to the standard
NNLO weak-coupling expansion pressure, illustrating how the RGOPT reduces the residual renormalization scale dependence.
Finally, in section V we consider a further modification from a physically more realistic quark mass spectrum, including the
genuine (current) strange quark mass ($N_f=2^*+1^*$), and we present our main results. We conclude in section VI. \\
A number of technical details are treated in several Appendices,
notably our nontrivial generalization to the case of arbitrary quark masses of the basic calculations of NNLO massive contributions of \cite{Kurkela:2009gj} is presented in Appendix C. 
\section{NNLO massive quark matter $N_f=2+1^*$ pressure}
We briefly review here the main calculation steps and explicit results for the NNLO weak-coupling expansion 
of the cold quark matter pressure with massive quarks, originally obtained in \cite{Kurkela:2009gj},
those results being a basic ingredient for our main purpose with the subsequent RGOPT construction.
As already mentioned, we fully reproduce all the analytical
expressions obtained in the above quoted work
(up to a few typos, easily identifiable from comparing the expressions in \cite{WebSiteRomatchke} used in \cite{Kurkela:2009gj}). Accordingly, the sequel of this 
Section contains almost verbatim expressions, adopting the same notations for clarity,
and at a level of details needed to make clear the modifications induced upon
considering the RGOPT construction in Sec.\ref{sec:RGOPT} below. 

\subsection{ Massive cold quark matter pressure}
\subsubsection{NLO pressure}
\begin{figure}[h!]
    \epsfig{file=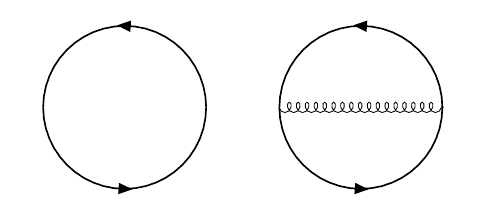,width=0.3\linewidth,angle=0}
    \caption{Contributions at LO and NLO to the QCD pressure.\label{Fig:GraphNLO}}
    \end{figure}
For symmetric quark matter (i.e. with chemical potentials $\mu_u=\mu_d=\mu_s\equiv \mu \equiv \mu_B/3$), 
the LO and NLO contributions in Fig. \ref{Fig:GraphNLO} to the free energy $\Omega=-P V$ with $P$ the pressure, for one massive quark flavor and after renormalization, reads
\begin{equation}\label{eq:PressureNNLO}
P(m,\mu) = P_{\rm LO}(m,\mu) + P_{\rm NLO}(m,\mu) + P_{\rm NNLO}(m,\mu)
\end{equation}
where 
\begin{equation}\label{eq:PressureLOpart}
P_{\rm LO}(m,\mu) =
\Theta (\mu^2 - m^2) \frac{N_c}{12 \pi ^2}\left[\mu\ p_F\left(\mu^2-\frac{5}{2}m^2\right) +\frac{3}{2}m^4 \ln \left(\frac{\mu + p_F}{m}\right) \right]
\end{equation}
\begin{equation}\label{eq:PressureNLOpart}
\begin{aligned}
P_{\rm NLO}(m,\mu) = 
    &-\Theta(\mu^2-m^2)\frac{d_A g^2}{4(2\pi)^4}\left\{3\left[m^2 \ln\left(\frac{\mu+p_F}{m}\right)-\mu\ p_F\right]^2-2p_F^4\right\} \\
    &-\Theta(\mu^2-m^2)\frac{d_A g^2}{4(2\pi)^4}m^2 \left(4-6\ln\frac{m}{\Lambda}\right)\left[\mu\ p_F-m^2\ln\left(\frac{\mu+p_F}{m}\right)\right], \\
\end{aligned}
\end{equation}
where
$g\equiv g(\Lambda)$ is the 
$\msbar$-scheme renormalized QCD coupling at renormalization scale $\Lambda$,
$d_A\equiv N_c^2-1$, $m$ is at the moment an arbitrary (renormalized) quark mass, and $p_F=\sqrt{\mu^2-m^2}$ is the Fermi momentum. 
\subsubsection{NNLO pressure: perturbative three-loop contributions}
\begin{figure}[h!]
    \epsfig{file=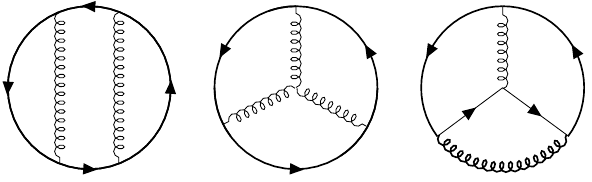,width=0.4\linewidth,angle=0}
    \caption{The 2GI contributions at NNLO.\label{Fig:GraphNNLO}}
    \end{figure}
\begin{figure}[h!]
    \centering
     \epsfig{file=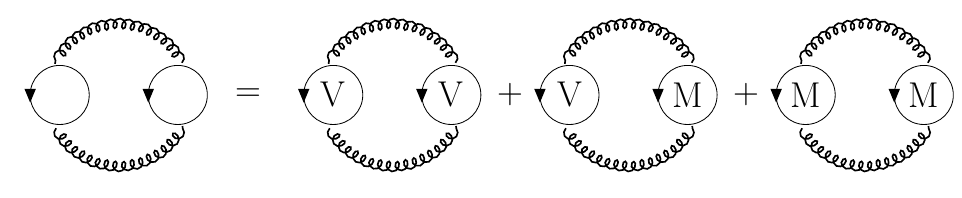,width=0.6\linewidth,angle=0}
    \caption{The plasmon contribution and its decomposition into three different part due to the gluon polarization tensor own decomposition into vacuum (V) and matter (M) part. The last one is absorbed in the ring contribution in Fig. \ref{Fig:Ring}.}
    \label{Fig:VV_VM_MM}
\end{figure}
\begin{figure}[h!] 
    \epsfig{file=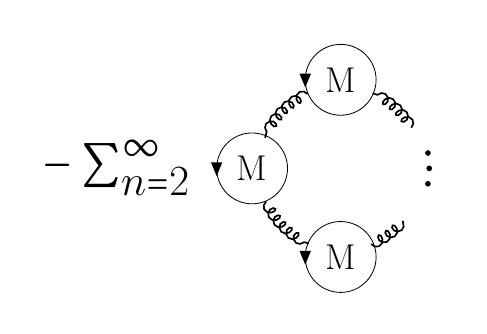,width=0.3\linewidth,angle=0}
    \caption{The ring sum contributing at $\mathcal{O}(\alpha_s^2)$.}  \label{Fig:Ring}
\end{figure}

At NNLO, two sets of diagrams can be identified\cite{pQCDmu4L,Kurkela:2009gj}, the IR and UV finite (after renormalization\footnote{The
mass and coupling renormalizations up to NNLO are briefly reviewed in Appendix A.}): 
the two-gluon-irreducible (2GI) contribution, see Fig. \ref{Fig:GraphNNLO}, and the IR divergent plasmon contribution appearing in Fig. \ref{Fig:VV_VM_MM}. The latter involving two independent quark loops, therefore, even the single-flavor contribution to the pressure now intrinsically depends on the full quark mass spectrum of the theory. The matter-matter part of the plasmon develops an IR divergence that is taken care of by resumming the set of diagrams, the ``ring'' resummation, shown in Fig. \ref{Fig:Ring}. As is well known\cite{soft1}, this resummation leads to a contribution $\sim \as^2 \ln \as\sim g^4\ln g$ (in the massless quark approximation) that breaks the naive perturbative expansion. 
As it will be clear in the explicit expressions below (see Eq.(\ref{eq:Ring2p1Star}), considering a non-vanishing quark mass involves similar $g^4\ln g$ terms
plus additional complicated quark mass dependence.
The pure vacuum VV contribution in Fig. \ref{Fig:VV_VM_MM} being independent of the chemical potential, is often discarded from the pressure contribution 
(we will however return to this point in Sec. \ref{sec:RGOPT} below), 
while the vacuum-matter VM diagram contributes at $\mathcal{O}(\as^2)$. The 2GI and the VM contributions for massive quarks were first calculated in \cite{Kurkela:2009gj}, more precisely for two massless flavors and one massive flavor (referred here as $N_f=2+1^*$).
The calculation can be conveniently organized in the form of 
separate 1-, 2-, and 3-cut contributions: in short, for a given graph this corresponds to the number (ranging from zero to the number of loops)
of fermionic lines being cut, i.e. with corresponding propagator put on-shell and remaining three-dimensional integration with the appropriate weight
$-\Theta(\mu-E_p)/(2E_p)$, where $E_p=\sqrt{p^2+m^2}$. Then, the remnant factors and extra integrations are to be performed for their corresponding vacuum ($\mu=0$) expressions. Note that the zero-cut, pure vacuum contribution, is again 
usually discarded in the cold quark matter literature. 
We refer to \cite{Kurkela:2009gj} for more details on these cutting rules (see also \cite{cutrules}). Here we simply 
display their final results, that we have fully reproduced upon following the same procedure.
\begin{equation}\label{eq:PhNNLO}
P_{\rm 2GI+VM}^{N_f=2+1^*} = \Theta(\mu^2-m^2)\frac{g^4}{(4\pi)^4}\mathcal{M}_{3}^{N_f=2+1^*} 
\equiv \Theta(\mu^2-m^2) \frac{g^4}{(4\pi)^4}\left( \mathcal{M}_{3}^{\rm 1c}+
\mathcal{M}_{3}^{\rm 2c}+
\mathcal{M}_{3}^{\rm 3c}\right) .
\end{equation}
 Accordingly, the renormalized 1-cut contribution reads: 
\begin{equation}\label{1cut}
\begin{split}
    \mathcal{M}_{3}^{\rm 1c}=&\frac{d_A\mu^4}{(2\pi)^2}\Bigg\{ \Bigg(-C_A\Big[\left(22\ln\frac{\Lambda}{m}+\frac{185}{3}\right)\ln\frac{\Lambda}{m}+\frac{1111}{24}-\frac{4\pi^2}{3}+4\pi^2\ln2-6\zeta(3)\Big]\\
    &-C_F\Big[3\left(12\ln\frac{\Lambda}{m}+5\right)\ln\frac{\Lambda}{m}+\frac{313}{8}+\frac{35\pi^2}{6}-8\pi^2\ln2+12\zeta(3)\Big]+N_f\Big[\frac{2}{3}\left(6\ln\frac{\Lambda}{m}+13\right)\ln\frac{\Lambda}{m}+\frac{71}{12}+\frac{2\pi^2}{3}\Big]\\
    &+6-2\pi^2\Bigg)\hat{m}^2 z+4\hat{m}^2C_F\Big[3\left(3\ln\frac{\Lambda}{m}+4\right)\ln\frac{\Lambda}{m}+4\Big](\hat{u}-z)\Bigg\},
\end{split}
\end{equation}
where $ \hat{m}= m/\mu$ , $\hat{u} = p_F/\mu = \sqrt{1-\hat{m}^2}$ and $z=\hat{u}-\hat{m}^2 \ln{\frac{1+\hat{u}}{\hat{m}}}$, $C_A=N_c$ and $C_F=\frac{N_c^2-1}{2N_c}$.
Next, the renormalized 2-cut contribution is
\begin{equation}\label{2cut}
    \begin{split}
        \frac{\mathcal{M}_{3}^{\rm 2c}}{(4\pi)^2}=&d_A\Bigg\{C_A\Big(-\frac{16}{9}I_1^2+\frac{62}{9}m^2I_2+\frac{5}{3}I_{\rm 1c}-\frac{10}{3}m^2I_{\rm 2c}+I_{10}-\frac{22}{3}[I_1^2-2m^2I_2]\ln\frac{\Lambda}{m}\Big)\\
        &+C_F\Big(I_{11}+24m^2[I_2-I_{\rm 1b}I_1+2m^2I_{\rm 2b}+2m^2I_8]\ln\frac{\Lambda}{m}\Big)+N_f\Big(\frac{10}{9}I_1^2-\frac{20}{9}m^2I_2-\frac{2}{3}I_{1c}+\frac{4}{3}m^2I_{\rm 2c}\\
        &+\big[\frac{4}{3}I_1^2-\frac{8}{3}m^2I_2\big]\ln\frac{\Lambda}{m}\Big)-\frac{2}{3}I_{12}\Bigg\},
    \end{split}
\end{equation}
and finally the 3-cut contribution (which is directly UV finite) reads 
\begin{equation}\label{3cut}
\begin{split}
    \frac{\mathcal{M}_{3}^{\rm 3c}}{(4\pi)^4}=&-d_A\Bigg\{C_A\Big[2I_1 I_2-4(I_5+I_7-2m^4I_6)\Big]+C_F\Big[2I_1^2I_{\rm 1b}-4I_1 I_2-8m^2I_1 I_{\rm 2b}+8m^2I_3+8m^4I_{\rm 3b}-2I_4\\
    &+8(I_5+I_7-2m^4 I_6)-8m^2I_1 I_8+8m^4 I_9\Big]\Bigg\},
    \end{split}
\end{equation}
where the basic integral $I_1(\hat m)$-$I_{12}(\hat m)$ for the quark mass dependence 
were originally defined in Appendix D of \cite{Kurkela:2009gj}. For self-containedness we 
reproduce those definitions in Appendix B, discussing in addition a number of useful relations
among these quantities.
Notice that the 1-cut contribution in Eq.(\ref{1cut}) has a fully analytical expression, while the basic integrals entering the 2-cut and 3-cut contributions, apart from $I_1, I_2, I_8$ that have
relatively simple analytical expressions, 
can only be evaluated numerically, some of these involving up to six-dimensional integration over appropriate 
momenta and angular variables. Accordingly, in Ref.\cite{Kurkela:2009gj} the integrals $I_k$
were evaluated numerically to good accuracy and subsequently fitted in the relevant range $0\le \hat m \le 1$, 
resulting into the more compact and convenient expression for the $N_f=2+1^*$ massive NNLO contribution Eq.(\ref{eq:PhNNLO}):
\begin{equation}\label{eq:PressureM3}
\begin{aligned}
  &  \mathcal{M}_{3}^{\small{N_f=2+1^*}} =\frac{d_A \mu^4}{2\pi^2}\Bigg\{-\hat{m}^2[(11C_A-2N_f)z+18C_F(2z-\hat{u})](L_m)^2 \\
    &-\frac{1}{3}\bigg[C_A\big(22\hat{u}^4-\frac{185}{2}z \hat{m}^2-33z^2\big) +\frac{9C_F}{2}(16\hat{m}^2\hat{u}(1-\hat{u})-3(7\hat{m}^2-8\hat{u})z-24z^2)\\
    &-N_f(4\hat{u}^4-13z\hat{m}^2-6z^2)\bigg]  L_m +C_A\left(-\frac{11}{3}\ln{\frac{\hat{m}}{2}}-\frac{71}{9}+G_1(\hat{m})\right)+C_F\left(\frac{17}{4}+G_2(\hat{m})\right)\\
    &+N_f\left(\frac{2}{3}\ln{\frac{\hat{m}}{2}}+\frac{11}{9}+G_3(\hat{m})\right)+G_4(\hat{m})\Bigg\}, \\
    \end{aligned}
\end{equation}
with $L_m \equiv \ln(m/\Lambda)$ and the $G_i(\hat m)$ fitting functions, defined such that $G_i(\hat m)\to 0$ for $\hat m \to 0$, 
were given explicitly in Eqs.(41)-(46) of Ref.\cite{Kurkela:2009gj}. Note that the 1-cut contribution in Eq.(\ref{1cut}) has a straightforward limit, $\mathcal{M}_3^{1c}\to 0$
for $m\to 0$, while the 2-cut and 3-cut contributions have far less trivial $m\to 0$ behavior,
with apparently uncanceled IR divergent $\ln (m)$ terms within Eqs.(\ref{2cut}), (\ref{3cut}), due to the lack of
analytical result for the $I_k$ integrals in these contributions. 
Actually, the necessary cancellations of such IR divergent 
terms, together with the known\cite{pQCDmu4L} $m\to 0 $ limit of Eq.(\ref{eq:PressureM3}), 
and explicit expressions of $I_k(m)$ (see Appendix B) provide useful constraints to determine the fitting
functions $G_i(x)$. At this stage, however, having
identified a mismatch between the function $G_2(\hat m)$ and its equivalent expressions
in terms of basic $I_k$ integrals (for which we found excellent agreement with 
available results \cite{WebSiteRomatchke} used in \cite{Kurkela:2009gj}), 
we have performed more independent crosschecks, obtaining a new, slightly more precise, determination of the above fitting functions 
$G_i,\ i\in\{1,2,3,4\}$. Our results are

\begin{equation}
    \begin{aligned}
          G_1(\hat{m}) = & \ 32 \pi^4 \hat{m}^2 \Big(0.001715 - 0.000339\,\hat{u}+ 0.002818\,\hat{u}^2 - 0.002282\, \hat{u}^3\\
   &+ 0.005854\ln\hat{m} - 0.018427\,\hat{m}^2 \ln^3\hat{m}+0.000444\, \hat{m}^2  \ln\left(\frac{1+\hat{u}}{\hat{m}}\right)\Big)\\
   G_2(\hat{m}) = &\ 32 \pi^4 \hat{m}^2 \Big(-0.001363+0.000401\, \hat{u}+0.003454\,\hat{u}^{2} - 0.002983\hat{u}^{3}+0.021502 \hat{u}^{4}\\
    &+ 0.017914\ln\hat{m}- 0.032789\,\hat{m}^2\ln^2\hat{m}+0.002067\,\hat{m}^2 \ln\left(\frac{1+\hat{u}}{\hat{m}}\right)\Big) \\ 
   G_3(\hat{m}) = & \ 32 \pi^4 \hat{m}^2 \Big(-0.000244 - 0.002192 \,\hat{u}+ 0.000086\,\hat{u}^2+0.001895\,\hat{u}^3\\
   &+0.000054\ln\hat{m}+0.000521\ln^2\hat{m}+ 0.002176 \,\hat{m}^2\ln\left(\frac{1+\hat{u}}{\hat{m}}\right) \Big)\\
   G_4(\hat{m}) = &\ 32 \pi^4 \hat{m}^2 \Big( -0.0020405\,\hat{u} +0.0003254\,\hat{u}^3+ 0.0001777\,\hat{u}^4+0.0002580\hat{u}^5-0.0003811\ln\hat{m} \\
   &- 0.0003289\ln^2\mh+0.0005292\hat{m}^2\ln\hat{m}+0.0004012\mh^2 \ln^2\mh+  0.0020462\,\hat{m}^2\ln\left(\frac{1+\hat{u}}{\hat{m}}\right)\Big) \\
    \end{aligned}\label{eq:GiFunc}
\end{equation}
We obtain excellent numerical agreement with \cite{Kurkela:2009gj} for the fitting functions 
$G_1$, $G_3$, $G_4$ (although the analytical expressions appear somewhat different,
due to slightly different functional fitting choices), but 
sizable differences in $G_2(\hat m)$ for intermediate
and large $\hat m $ values. This is discussed in more details in Appendix B, where
a comparison is shown in Fig. \ref{Fig:Gi_Comparison}. Although in Fig. \ref{Fig:Gi_Comparison} the $G_2$ discrepancies appear important, note that the generic massive result Eq.(\ref{eq:PressureM3}) 
is only used for the strange quark mass $m_s(\Lambda)$
contributions within the standard NNLO $N_f=2+1^*$ pressure. Accordingly, accounting for its running, $ m_s(\Lambda)/\mu \le 0.15$ roughly within the range $\mu \ge 0.7$ GeV (below which $\as(\Lambda\sim \mu)$ becomes anyway too large for perturbative results to be much reliable), namely one always has $m_s(\Lambda)/\mu \ll 1$. 
In this range our obtained differences in $G_2(\hat m)$ are very moderate, as seen in Fig. \ref{Fig:Gi_Comparison}. Moreover, the terms involving $G_1(\hat m)$ 
numerically dominates the contributions to Eq.(\ref{eq:PressureM3}), such that overall the updated $G_2$ expression in Eq.(\ref{eq:GiFunc}) has a benign impact on the complete $N_f=2+1^*$ NNLO pressure as compared to \cite{Kurkela:2009gj}. 
This will be illustrated in Fig. \ref{Fig:RelativeErrorPQCDnewGi} after completing 
the remaining needed contributions to the NNLO pressure below.
\subsubsection{Plasmon ring resummed contribution}
The plasmon (ring) contribution to the free energy $\Omega\equiv -P V$, that corresponds to resummed quark loops, for the relevant
case of $N_l$ massless quarks and one massive quark, reads (keeping strictly 
the same notations as in \cite{Kurkela:2009gj}):
\begin{eqnarray}\label{eq:Ring2p1Star}
&& \Omega^{N_f=2+1^*}_{\rm Ring}=\frac{d_A\ g^4}{512\pi^6}
\Bigg\{ (\Vec \mu^2)^2 \left[ \left(2\ln\left(\frac{g}{4\pi}\right)-\frac{1}{2}\right)+\frac{1}{2}\left(-\frac{19}{3}+\frac{2\pi^2}{3}+\frac{I_{15}(\Vec{\mu})}{(\Vec \mu^2)^2} +\frac{16}{3}(1-\ln2)\ln(2)+I_{16}(\hat{m},\Vec{\mu}^2)\right)\right] \nonumber\\
&&  +  2\mu^2 \sum_{i=1}^{N_l=2} \mu_i^2 \left[ I_{14}\left( 2\ln\left(\frac{g}{4\pi} \right)-\frac{1}{2}\right) +\frac{1}{2}\left(I_{17}(\hat{m},\mu_i)+\frac{16}{3}(1-\ln2)\ln(2)\,I_{18}+I_{19}(\hat{m},\Vec{\mu}^2) \right)\right]
\nonumber\\
&& 
+\mu^4 \left [ I_{13}\left(2\ln\left(\frac{g}{4\pi}\right)-\frac{1}{2}\right)+\frac{1}{2}\left(I_{20}(\hat{m})+\frac{16}{3}(1-\ln2)\ln(2)I_{21}+I_{22}(\hat{m},\Vec{\mu}^2) \right)
\right]
\Bigg\} ,\ \ \vec{\mu}=\(\mu_1,\mu_2\)
\end{eqnarray}
where all basic integrals $I_{13}-I_{22}$ were given explicitly in Appendix D of \cite{Kurkela:2009gj} and will not be repeated here.
\subsubsection{NNLO pressure: additional mixed VM graph}
Finally, upon considering the summed contributions of $N_l$ massless (u,d) quarks and one massive (strange) quark,
there is an additional vacuum-matter (VM) term, such that the matter quark loop is massless, while the vacuum term 
is the difference between massive and massless contributions. 
An approximate but accurate expression reads\cite{Kurkela:2009gj}
\begin{equation}\label{eq:VMx}
\Omega^{N_f=2+1^*}_{{\rm VM} ,x} = d_A \frac{g^4}{(4\pi)^2} \frac{m^4}{12\pi^2} \sum_{i=1}^{N_l=2} 
I_x \left(\frac{\mu_i}{m+\mu_i}\right),
\end{equation}
where $N_l$ stands for the number of massless flavors and $I_x$ is a well-defined integral,
that is properly approximated by the following expression
\begin{equation}
I_x(t) = -3t^4(1-\ln t) \left[\frac{0.83}{(1-t)^2} +\frac{0.06}{(1-t)}-0.056
+ \frac{\ln(1- t)}{t(1-t)^2}( 1.005-0.272t (1-t) +0.154 t(1-t)^2) \right].
\end{equation}
\subsubsection{Remaining contribution from the massless quarks}
 The last quantity required is the contribution from the massless part entering at LO, NLO,
 and from 2GI plus VM graphs (NB the massless quark contributions within the resummed ring diagram is already
 accounted within Eq.(\ref{eq:Ring2p1Star}), as is explicit): 
Its well-known result for symmetric quark matter (see e.g. \cite{pQCDmu4L}) reads, as given in Eq.(23) of \cite{Kurkela:2009gj}
\be\label{PNLOm0}
  P^{m=0}= \frac{1}{4\pi^2}\sum_{i=1}^{N_l=2}\mu_i^4\left\{\frac{N_c}{3}-d_A\frac{g^2}{(4\pi)^2}+d_A\frac{g^4}{(4\pi)^4}\left[\frac{4}{3}\(N_f-\frac{11\,C_A}{2}\)\ln\frac{\Lambda}{2\mu_i}-\frac{142}{9}C_A+\frac{17}{2}C_F+\frac{22}{9}N_f\right]\right\}.
\ee
\subsection{Numerical results and comparisons: $N_f=2+1^*$ pressure} 
We illustrate at this stage the resulting standard NNLO perturbative quark matter pressure, $N_f=2+1^*$, obtained from combining all the previous contributions.
To recap, the total NNLO quark matter pressure, for two massless (u,d) and one massive (strange) quark is
\begin{equation}\label{eq:P2+1NNLO}
\begin{aligned}
\mathcal{P}_{\rm QCD}^{N_f=2+1^*}(\mcm,\mu)=&  P^{m=0}(\mu) +P_{\rm LO}(\mcm,\mu)+P_{\rm NLO}(\mcm,\mu)+ P^{\rm N_f=2+1^*}_{\rm 2GI, VM}(\mcm,\mu)\\
&-\Omega^{N_f=2+1^*}_{\rm Ring}(\mcm,\mu)-\Omega^{\rm N_f=2+1^*}_{\rm VM,x}(\mcm,\mu)
\end{aligned}
\end{equation}
obtained respectively from  \Eq{\ref{eq:PressureLOpart}},(\ref{eq:PressureNLOpart}),(\ref{eq:PressureM3}),(\ref{eq:Ring2p1Star}),(\ref{eq:VMx}),(\ref{PNLOm0}).
\subsubsection{Running mass and coupling prescriptions}
\begin{figure}[h!]
    \hspace{-3cm}
    \begin{subfigure}[h]{0.4\textwidth}
        \epsfig{file=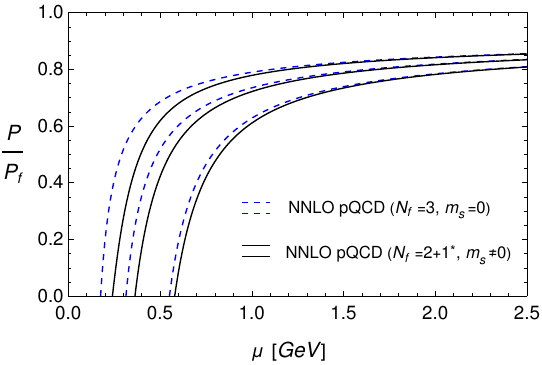,width=1.28\linewidth,angle=0}

    \end{subfigure}
    \hspace{3cm}
    \begin{subfigure}[h]{0.4\textwidth}
         \epsfig{file=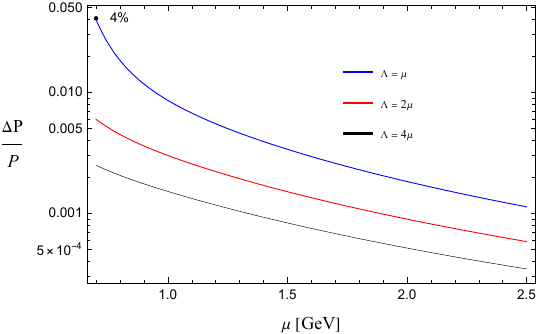,width=1.35\linewidth,angle=0}
    \end{subfigure}
     \caption{Left: the normalized pressure 
     $\mathcal{P}_{\rm QCD,NNLO}^{N_f=2+1^*}/\mathcal{P}_f$ obtained 
     from our reevaluated $G_i$ functions, Eq.(\ref{eq:GiFunc}), with
     residual scale dependence $\mu \le \Lambda \le 4\mu$. Right:
     relative difference with respect to KRV 2010 \cite{Kurkela:2009gj} for
     three different renormalization scales as indicated in the legend,
     with $\Delta P\equiv 
     P(\mathrm{KRV \;2010})-P(\mathrm{this \; work})$, and $P$
     refers to $\mathcal{P}_{\rm QCD,NNLO}^{N_f=2+1^*}$ in Eq.(\ref{eq:P2+1NNLO}).
     }
     \label{Fig:RelativeErrorPQCDnewGi}
\end{figure}

For the running coupling $g^2(\Lambda)\equiv 4\pi\as(\Lambda)$, 
we use the exact NLO one, obtained for a given renormalization 
scale $\Lambda$ from solving 
\begin{equation}\label{eq:RunningNNLO}
\Lambda_{\rm \overline{MS}}= \Lambda e^{-\frac{1}{2b_0 g^2}}\left(\frac{b_0 g^2}{1+\frac{b_1}{b_0} g^2} \right)^{-\frac{b_1}{2b_0^2}}\ 
\end{equation}
with beta-function coefficients defined
in our normalization in Appendix A, and fixing
$\Lambda_{\rm \overline{MS}}=330\ $MeV\cite{PDG2018} so that $\alpha_s(\Lambda = 1.5 \, {\rm GeV})\simeq 0.326$ \cite{Bazavov:2012ka}.
Using an higher order running coupling would hardly give any visible 
differences in our numerical results for the relevant $\mu$ range considered.
For the strange quark, since mass renormalization is only needed at NLO in the present case,
we use the NLO running mass expression, given in our normalization 
as:
\be\label{msrun}
\mcm(\Lambda)=m_s(\Lambda_0)\(\frac{g^2(\Lambda)}{g^2(\Lambda_0)}\)^{\frac{\gamma_0}{2b_0}}\(\frac{1+\frac{b_1}{b_0}\,g^2(\Lambda)}{1+\frac{b_1}{b_0}\,g^2 (\Lambda_0)}\)^{\frac{\gamma_1}{2b_1}-\frac{\gamma_0}{2b_0}}
\ee
where $m_s(\Lambda_0=\mathrm{2 GeV}) \simeq 93.5$ MeV\cite{PDG2018}.
Similarly to the running coupling, accounting for higher orders in the running strange quark mass does not lead to evident differences in the pressure. \\
The NNLO pressure resulting from Eq.(\ref{eq:P2+1NNLO}), normalized to the free pressure $\mathcal{P}_f =N_c N_f\, \mu^4 /(12\pi^2)$, is shown in Fig.~\ref{Fig:RelativeErrorPQCDnewGi} (left), also
compared to the massless quark $N_f=3$ pressure.
The residual renormalization scale dependence is illustrated from taking $\mu \le \Lambda \le 4\mu$ as is conventionally done in the literature. Note that the scale dependence of both $N_f=3$ and $N_f=2+1^*$ pQCD pressures in Fig. \ref{Fig:PressureNNLO}
appears slightly reduced compared to \cite{Kurkela:2009gj}, due to the exact NLO running coupling Eq.(\ref{eq:RunningNNLO}) 
producing slightly smaller scale variations for very low scale values, compared to
the more standard running coupling given as an expansion in $\ln(\Lambda)^{-1}$\cite{PDG2018}.
We also illustrate in Fig. \ref{Fig:RelativeErrorPQCDnewGi} (right) 
the modifications induced from our updated $G_i$ fitting functions 
(notably $G_2$): comparing the two pressures obtained respectively from the previous\cite{Kurkela:2009gj} or with our new $G_i$ functions 
in Eq.(\ref{eq:GiFunc}), we observe that the $G_2(\hat m)$ 
difference we obtained has a very moderate effect on the standard NNLO pressure 
values for $\mcm \ne 0$: in the extreme case, for the scale $\Lambda=\mu$ and $\mu=0.7$ GeV,
where $\as\approx 0.58$ is already very large, the difference is about $-4\%$ (noting that
the updated $G_2$ gives a {\em lower} pressure for a given $\mu$ value). 
In contrast, the differences in $G_2$ will have more impact for our RGOPT construction, as the latter involves medium-dressed quark masses of order $ \sim g\, \mu \gg m_s$, as
it will be developed in the next Section.  
\subsubsection{Pocket formula for pQCD $N_f=2+1^*$}
Finally, for completeness, we provide a compact fitting function
giving a good approximation of the full NNLO quark matter $N_f=2+1^*$ pressure, in the spirit of a similar approximation
given in Ref.\cite{prevN2LOfit}:
\ba\label{eq:PocketFormulaPQCD}
&&\frac{\mathcal{P}_{\rm QCD}^{N_f=2+1^*}(\mu,\Lambda=X\, \mu)}{\mathcal{P}_{f}(\mu,N_f=3)}= \(c_1 + c_2  X^{\nu_3}\) - \frac{ d_1   (3 \Tilde{\mu})^{\alpha_1}X^{\nu_1}}{(3 \Tilde{\mu} - d_2\, X^{-\nu_2}) },\ \ \  \Tilde{\mu}=\mu/{\rm GeV}, \nn
&& c_1=0.830189 , c_2=0.505545 , d_1=0.438396, d_2=1.165107, \alpha_1=1.014939, \nn
&& \nu_1=0.670277, \nu_2=0.899925, \nu_3=0.632526.
\ea

\section{RGOPT pressure at next-to-next-to-leading order for $N_f=3^*$}\label{sec:RGOPT}
Although the quark sector of dense QCD does not exhibit proper IR divergences like for the soft gluon modes, 
it is nevertheless highly desirable to resum well-defined (RG-induced) 
higher orders as well
in this sector, 
incorporating additional information beyond the strictly perturbative expansion.
Accordingly, our main aim is to possibly further reduce in particular the residual renormalization
scale dependence, which hampers to some extent
the standard cold quark matter pressure perturbative expansion. 

Quite similarly to the SPT\cite{spt1,spt3} or HTLpt\cite{HTLpt} framework, the RGOPT implies as 
a first step to modify the original QCD Lagrangian by adding and subtracting a Gaussian (quark) mass term, with one of the contribution being treated as an interaction. Although the
specific mass dependence is at first the one dictated by the standard massive perturbative calculations, 
this mass is subsequently treated as a variational parameter, $\overline m$, not to be confused with 
the physical quark masses in the present context.
More precisely, as will be derived in more details below, $\overline m$ is
determined\cite{RGOPTals,rgopt_cold} from a self-consistent mass gap upon imposing the pressure $P(\overline m)$
to satisfy the RG equation at a given order of the modified expansion. 
This gives to $\overline m$ the properties of a medium-dressed quark mass, $\overline m (g,\mu\cdots)$, 
but embedding an RG-dictated all order coupling dependence.
Accordingly, at successive orders of the modified expansion,
the resulting pressure $P(\overline m(g,\mu,\cdots))$ gives a sequence of improved approximations to the initially massless quark limit of the theory.

To simplify, we first implement the RGOPT for 
degenerate (dressed) quark masses, i.e. the $N_f$ quark flavors are treated similarly, obtaining a common
RG-dressed medium mass $\overline{m}(\mu)$, while the physical quark masses are neglected. Then we consider the more precise non-degenerate case, incorporating in this framework the extra contribution from the genuine strange quark (current) mass. 
Concretely, we first have to modify the previously considered $N_f=2+1^*$ 
pressure to implement $N_f=3^*$ degenerate massive quarks, followed by a generalization to unequal masses.  

\subsection{From $N_f=2+1^*$ to $N_f=3^*$ massive quarks}
Upon examination of the relevant contributions, including the cancellations of UV divergences and renormalization terms, going from $N_f=2+1^*$ to $N_f=3^*$ massive quarks with an identical mass can be incorporated upon following the simple prescriptions:
\begin{equation}
\begin{array}{ccl}\label{M1c2cmod}
 \mathcal{M}_{3}^{\rm 1c}& \to &\mathcal{M}_{3}^{\rm 1c}+
(N_f-1) 2(3-\pi^2) \hat m^2 z , \\
 \mathcal{M}_{3}^{\rm 2c} &\to &\mathcal{M}_{3}^{\rm 2c}-(N_f-1)  \frac{2}{3}I_{12} ,\\
 \mathcal{M}_{3}^{\rm 3c}&\to & \mathcal{M}_{3}^{\rm 3c}
\end{array} \Bigg\}\Rightarrow G_4(\hat{m}) \to N_f G_4(\hat{m}), \ \ {\rm for}\ N_f=2+1^*\to 3^*
\end{equation}
which accordingly modifies $\mathcal{M}_3^{N_f=2+1^*}$ to $\mathcal{M}_3^{N_f=3^*}$ in Eq.(\ref{eq:PressureM3})
with the same $G_i$ functions in Eq.(\ref{eq:GiFunc}), and then
\be\label{eq:P32GIVM}
P^{\rm N_f=3^*}_{\rm 2GI, VM}(m,\mu) = \frac{g^4}{(4\pi)^3} \mathcal{M}_3^{N_f=3^*}.
\ee
On top of Eq.(\ref{M1c2cmod}), an extra overall
factor of $N_f$ multiplies the contributions in \Eq{\ref{eq:PressureNNLO}}, 
so that explicitly:
\begin{equation}\label{eq:P3NNLO}
\mathcal{P}_{\rm NNLO}^{N_f=3^*}(m,\mu)= N_f\(P_{\rm LO}(m,\mu)+P_{\rm NLO}(m,\mu)+ P^{\rm N_f=3^*}_{\rm 2GI, VM}(m,\mu)\)-\Omega^{N_f=3^*}_{\rm Ring}(m,\mu)\; .
\end{equation}
Note that the 
vacuum-matter term
$\Omega_{{\rm VM},x}^{N_F=3^*}$ appearing in Eq.(\ref{eq:VMx}) is now zero since $N_l=0$. Concerning next the ring contributions, care must be taken to avoid double counting
contributions and other inconsistencies when generalizing the results from $N_f-1$ massless
quarks plus a single massive quark, to $N_f$ massive quarks. The final expression reads
\begin{equation}\label{eq:Ring3star}
\Omega_{\rm Ring}^{N_f=3^*}=\frac{N_f^2 d_A\ g^4\, \mu^4}{512\pi^6}\left[I_{13}\left(2\ln\left(\frac{g}{4\pi}\right)-\frac{1}{2}\right)+\frac{1}{2}\left(I_{20}+\frac{16}{3}(1-\ln2)\ln(2)I_{21}+I_{23} \right) \right],
\end{equation}
where the overall factor $N_f^2$ accounts for all degrees of freedom. Note that for this purpose ($N_f=3^*$) one has to define a new integral, $I_{23}$, generalizing $I_{22}$ of Eq.(\ref{eq:Ring2p1Star}) to three massive quarks with the same mass and chemical potential. The definition of $I_{23}$ is 
\begin{equation}
\begin{split}
    I_{23}(m,\mu)=&\frac{16}{\pi}\int_0^{\frac{\pi}{2}}d\phi \sin{\phi}^2\left(F_h^2(0,\phi,\hat{m})\ln(N_f F_h(0,\phi,\hat{m}))+\frac{1}{2}G_h^2(0,\phi,\hat{m})\ln(N_f G_h(0,\phi,\hat{m}))\right)\\    =&2\,I_{13}(m,\mu)\ln N_f+\delta(1-\hat{m}^2)+\hat{m}^2\left(1.4000-2.2193\hat{u}\right)+\hat{m}^4\left(-2.4866+2.8318\hat{u}\right)+\hat{m}^6\left(1.0866-0.5860\hat{u}\right),
    \end{split}
\end{equation}
where $\delta\approx -0.85638321$ was defined in \cite{pQCDmu4L}, and
$F_h$, $G_h$ are the same functions defined in Eq.(A11) of \cite{Kurkela:2009gj}. 
The corrections from different chemical potentials appear in $I_{20}$ and $I_{23}$ but have been neglected since their relative contribution is numerically completely negligible. 

\subsection{RGOPT modifications to the pressure}
The next important modification with respect to the above
weak-coupling expansion pressure in Eq.(\ref{eq:PressureNNLO}), is to reintroduce 
pure vacuum contributions, originally present in the basic calculation for any massive theory. These are evidently
vanishing for massless quarks, and even when considering the quark mass dependence 
they are often justifiably neglected in the quark matter literature on grounds
that they depend solely on the mass $m$ and not on the in-medium $\mu$ scale.
However, since RG properties are essentially related to UV divergences,
and the latter are determined by the $T=\mu=0$ theory,
these vacuum contributions play a crucial role in the {\em massive} RG invariance properties, as will 
be recalled below.

Up to NNLO, the complete contributions to the pure vacuum (including the VV part of the plasmon)
pressure can be easily extracted from 
\cite{KNcond}, 
modifying the above defined NNLO quark matter pressure expressions 
in Eqs.(\ref{eq:PressureLOpart}), (\ref{eq:PressureNLOpart}), (\ref{eq:PressureM3}), 
as follows: 
\begin{equation}\label{eq:PLOrgopt}
P_{\rm LO}(m,\mu) \to P_{\rm LO}(m,\mu) -N_{c} \frac{m^4}{8 \pi ^2}\(\frac{3}{4}-L_{m}\)
\equiv P^{v+m}_{\rm LO}(m,\mu)
\end{equation}
\begin{equation}\label{eq:PNLOrgopt}
\begin{aligned}
P_{\rm NLO}(m,\mu) \to  P_{\rm NLO}(m,\mu) 
    &-\frac{d_A g^2}{4(2\pi)^4}m^4 \left(3L_m^2-4L_m+\frac{9}{4}\right) \equiv P^{v+m}_{\rm NLO}(m,\mu)
\end{aligned}
\end{equation}
\begin{equation}\label{eq:PNNLOrgopt}
    P_{\rm NNLO}(m,\mu) \to  P_{\rm NNLO}(m,\mu) + P^{v,d}_{\rm NNLO}(m,N_f,N_h,N_l) \equiv P^{v+m}_{\rm NNLO}(m,\mu)
\end{equation}
where $ L_m=\ln (m/\Lambda)$, and \footnote{In Eq.(\ref{eq:CoefNNLO}) 
we distinguish the contributions $\propto N_f\equiv N_h+N_l$, actually originating from
lowest (two-loop) order RG coefficients, from the genuine three-loop $N_l$, $N_h$
inequivalent contributions, originating from the VV contributions in Fig. \ref{Fig:VV_VM_MM}. This distinction makes more transparent 
the modifications implied when considering two different masses for the quarks, 
see Appendix \ref{AppendixC2}.}

\begin{equation}
    \begin{aligned}
    P^{v,d}_{\rm NNLO}(m,N_f,N_h,N_l)\equiv&\frac{g^4 m^4}{\pi^2(4\pi)^4}\left(a_{3,3}+a_{3,2}\ L_m + a_{3,1}\ L_m^2 + a_{3,0}\ L_m^3 \right)\\
    a_{3,3}=&\ -\frac{23\,821}{144} -\frac{22}{3} \zeta(4) - 
   \frac{8}{3}\zeta(2) (\log{2})^2 +\frac{4}{9}(\log{2})^4 + 
   \frac{32}{3}\ {\rm Li}_4(1/2) -6\ \zeta(3) \\
   &+\frac{13}{12} N_f + N_h \(\frac{367}{24} -\frac{28}{3}\ \zeta(3)\)+N_l\(\frac{45}{8}+\frac{8}{3}\zeta(3)\)\\
    a_{3,2}= &\ \frac{3817}{12} -4 \zeta(3)-
    \frac{13}{3} N_{f} -\frac{39}{2}N_h -\frac{15}{2}N_l \\
   a_{3,1}= &  \frac{1}{3} \(-807 + 26 N_f\) \\
   a_{3,0}= & \frac{4}{3} \(81 - 2 N_f\)    
   \end{aligned}\label{eq:CoefNNLO}
\end{equation}
where in the most general case $N_l$, $N_h$ are respectively the number of massless and massive quarks and $N_f\equiv N_h+N_l$. The ``d'' index at NNLO stands for \textit{diagonal} since we will later distinguish the contributions from different quark masses in the ``VV'' diagram, Fig. \ref{Fig:VV_VM_MM}, contained in \Eq{\ref{eq:CoefNNLO}}. 
For the degenerate case $N_f=3^*$, we always have $N_l=0$ such that $N_h=N_f$, thus 
omitting here the argument. The generalization to $N_f=2^*+1^*$ 
in the next section is more subtle and will be addressed later. \\ 
This gives for the resulting complete NNLO vacuum and medium pressure contribution:
\begin{equation}\label{Ph+v}
P^{v+m}(m,\mu) =P^{v+m}_{\rm LO}(m,\mu)+P^{v+m}_{\rm NLO}(m,\mu)+
P^{v+m}_{\rm NNLO}(m,\mu)
\end{equation}
In particular, it is easily seen from Eq.(\ref{eq:PLOrgopt}) and Eq.(\ref{eq:PressureLOpart}) 
that the $\ln (m)$ terms cancel via the recombination 
$\ln \frac{m}{\Lambda} +\ln \frac{\mu+p_f}{m} =\ln \frac{\mu+p_f}{\Lambda}$. 
Similarly, this cancellation also occurs at NLO. Now, note importantly that for arbitrary $m$
the pressure in Eq.(\ref{Ph+v}) is not perturbatively RG invariant, as it explicitly 
depends on the renormalization scale $\ln(m/\Lambda)$ at LO, and within the original expression
Eq.(\ref{Ph+v}) this contribution cannot be canceled from $g(\Lambda)$ or $m(\Lambda)$ dependence,
that would only affect higher orders. This is a known feature of any massive theories, actually
related to their vacuum energy anomalous dimension.
Accordingly, an important preliminary step of our construction is 
to obtain a (perturbatively) RG invariant (RGI) pressure,
which implies to adding an extra zero point energy contribution, $S(m,g)$, to the pressure\cite{KNcond,rgopt_phi4}:
\begin{align}
 P^{v+m}(m,\mu) \to P^{v+m}(m,\mu) -S(m,g) \equiv 
 P^{v+m}(m,\mu) -\frac{m^4}{g^2} \sum_{k\ge 0} 
 s_k \(g^2\)^k \equiv P^{v+m}_{\mathrm{RGI}}(m,\mu),
 \label{sub}
\end{align}
such that the latter combination is approximately scale-independent, 
at a given perturbative order. More precisely,
the coefficients $s_k$ are determined at successive orders by applying the {\em massive} (homogeneous) RG operator
\begin{align}
 \Lambda \frac{d}{d \Lambda}= \Lambda \frac{\partial}{\partial \Lambda} + \beta(g^2 )\frac{\partial}{\partial g^2 } - 
\gamma_m(g^2 ) m \frac{\partial}{\partial m} \;,
\label{RG}
\end{align}
on Eq.(\ref{sub}), requiring it to vanish up to
neglected higher order terms. 
In our conventions the RG functions $\beta(g^2)$ and $\gamma_m(g^2)$ (the anomalous mass dimension) 
are given by
\begin{equation}
 \beta\left(g^2\equiv 4\pi\alpha_s \right)=
 -2b_0 g^4 -2b_1 g^6 +\cdots  \;,
 \label{betaQCD}
 \end{equation}
 and
 \begin{equation}
 \gamma_m\left(g^2\right)=\gamma_0 g^2+\gamma_1 g^4+\cdots\;,
\end{equation}
with higher orders and explicit expressions collected in Appendix A.
Accordingly, one obtains \cite{rgopt_cold} 
\begin{equation}
 s_0=- N_c \left[(4\pi)^2 (b_0-2\gamma_0)\right]^{-1} \;,
 \end{equation}
 and other relevant higher order $s_k, k=1,\cdots,3$ coefficients 
 are listed in Appendix A.\\
Note that in a fully equivalent way, $S(m,g)$ may be obtained from 
applying Eq.(\ref{RG}) as 
\be\label{eq:vacEn}
\Lambda \frac{d}{d \Lambda} \left[ S(m,g) \right] \equiv \hat\Gamma^0(g) m^4 = 
m^4 \sum_{k\ge 0} \Gamma^0_k g^{2k}
\ee
where $\hat \Gamma^0(g)$ defines the vacuum energy anomalous dimension\cite{vacEn1,vacEn2,E0phi4}, similarly relevant for 
other massive theories, and adding an inhomogeneous contribution to the RG operator\footnote{Renormalization aspects related to the vacuum energy contributions
are briefly overviewed in Appendix A.}
in Eq.(\ref{RG}).
For QCD, $\hat \Gamma^0(g)$ has been determined up to five-loop order\cite{vacEn5l}. 
Notice that $S(m,g)$ only depends on the vacuum contributions and not on the
medium ones. By construction the $s_k$ coefficients, even though being of order $k$-loops, encode RG information from order $k+1$. \\

Having reintroduced vacuum contributions, at this stage our practical definition
of the physical pressure is the perturbatively RG-invariant one at the relevant order, Eq.(\ref{sub}).
Next, the RGOPT is implemented as the following steps\cite{RGOPTals,rgopt_cold}:\\
1) First, reshuffling the interaction terms in the QCD Lagrangian,
according to the replacements
\begin{equation}
 m\to m \left(1-\delta\right)^a, \;\;  g^2\to \delta g^2,
 \label{delta}
\end{equation}
where in the present context $m$ is a quark mass\footnote{A rather similar treatment 
of the gluon sector is possible, starting from the HTL 
gauge-invariant effective Lagrangian\cite{HTLbasic,HTLpt} properly describing a gluonic mass term,
but is beyond our present scope.}. Eq.(\ref{delta}) can be most conveniently performed directly within the renormalized RGI massive pressure, Eq.(\ref{sub}): $\delta$ is a new expansion parameter, interpolating between the massive but free theory ($\delta\to 0$), and the massless 
interacting original theory, $\delta\to 1$, and the exponent $a$ will be specified below. 
For $a=1$ Eq.(\ref{delta}) is equivalent to the 
more familiar and intuitive ``added and subtracted'' mass term prescription, typically adopted in SPT\cite{spt1,spt3} or also in HTL perturbation theory\cite{HTLptDense2L,HTLptDense3L}. Eq.(\ref{delta}) is consistent with standard
mass renormalization, i.e. it does not add any new type of counterterms as long as one treats
the mass counterterms within the modified $\delta$-expansion consistently with Eq.(\ref{delta}).
As mentioned previously, we stress that $m$ is an {\it arbitrary} mass at this stage, determined from the prescription specified below.\\
2) Next, one expands the pressure resulting from Eq.(\ref{delta}) to the 
perturbative order in $\delta$ consistent with the usual perturbative expansion, 
then setting the result (after expansion) to $\delta\to 1$ suitable for the massless theory.
Now importantly, RGOPT is based on observing that such a modified perturbative expansion 
may generally spoil the perturbative RGI properties of the original physical quantity 
(the pressure Eq.(\ref{sub}) in the case at hand):
therefore the exponent $a$ in Eq.(\ref{delta}) is instead fixed (uniquely)
such that RG invariance is preserved.
 Accordingly, applying the massless RG equation (i.e. with $m=0$ in \Eq{\ref{RG}}), to the LO RGOPT pressure, one obtains 
 the critical value\cite{RGOPTals,rgopt_cold,rgopt_hot} 
 \begin{equation}\label{acrit}
a= \frac{\gamma_{0}}{2b_0},
\end{equation}
 thus determined solely by the first order RG function coefficients.\\
3) Since step 2) leaves a remnant $m$-dependence at any finite $\delta$ orders, 
similarly to the OPT\cite{opt_pms} (or HTLpt\cite{HTLpt,HTLptDense2L}) prescriptions, 
$m$ may be determined from requiring stationarity 
or mass optimization (OPT), 
\begin{equation}
 \frac{\partial P^{\rm{RGOPT}}}{\partial m} \Big \vert_{\overline{m} } = 0 \;,
 \label{opt}
\end{equation}
giving a sequence of improved approximations at successive orders 
of the actually massless all order result.
Accordingly $\overline m(g,\mu)$ is 
the nontrivial solution of a mass gap equation,
akin to a medium-dressed mass. \\ 
 4) According to our construction, $a$ is fixed once and for all at LO from Eq.(\ref{acrit}),
 then used at higher orders of the $\delta$-expansion as a sensible way of comparing successive perturbative orders.
 Whereas at LO the RG equation is automatically fulfilled 
 due to Eq.(\ref{acrit}), at NLO and higher orders the 
resulting pressure no longer satisfies the RG Eq.(\ref{RG}), due to reshuffled mass
 dependence.   
 Therefore, an alternative\cite{RGOPTals} to the OPT Eq.(\ref{opt}) is rather to (re)impose the RG Eq.(\ref{RG}), including RG coefficients consistently at the same perturbative order. 
The resulting nontrivial solution gives a ``RG-dressed'' medium mass, which by construction is expected to better restore the 
RG invariance of 
 the pressure, as it includes 
 higher order RG contributions.
 Actually, since we aim to recover the originally 
 massless theory, it is more appropriate to determine $\overline m$ upon solving  
 the {\em reduced} RG equation:
\begin{equation}\label{RGred}
 \left[ \Lambda\frac{\partial}{\partial \Lambda}\mathcal{P}+ \beta(g^2)\frac{\partial}{\partial g^2}\mathcal{P} \right]_{\overline{m}_{RG} }= 0 \;.
\end{equation} 
Unfortunately, at NLO and higher orders, the exact solutions of either 
Eq.(\ref{opt}) or Eq.(\ref{RGred}) are becoming highly nonlinear in $m$, with multiple $\overline m$ solutions appearing, some being non-real valued. 
As we will examine below, this issue can be remedied by a simple renormalization scheme
change, affecting the perturbative coefficients in such a way that a real solution can be recovered.
Moreover, importantly Eq.(\ref{acrit}) also implies the compelling feature that at higher orders the asymptotic freedom (AF) behavior  
for $\overline m(g\to 0)$ is obtained for only one of the solutions\cite{RGOPTals}. \\
For convenience the main steps to implement the RGOPT at NNLO are summarized
in Table \ref{tab-algo}. Apart from recapitulating  prescriptions
in items 1)-4) above, it contains the additional required ingredients 
discussed in more details in Sec. IV, and generalization to $m_s \ne 0$ fully treated
in Section V. We anticipate that at NNLO we obtain
a uniquely defined RGOPT prescription with a $\overline m(g,\mu)$ solution compatible
with a screening mass behavior.
\section{RGOPT for cold quark matter}
\subsection{NLO RGOPT}
Before addressing the more involved results at NNLO, we briefly
summarize the RGOPT results that were obtained at NLO\cite{rgopt_cold} considering three flavors of quarks,  corresponding to 
the graph in Fig. \ref{Fig:GraphNLO}. With the genuine mass
of the quarks being zero we end up with the chemical equilibrium condition $\mu_s=\mu_u=\mu_d\equiv \mu=
\mu_B/3$. At this order there are still no gluons 
in the medium as they formally enter at the NNLO through the Ring diagrams.
Starting from the known perturbative LO and NLO pressures in Eqs.(\ref{eq:PressureLOpart})(\ref{eq:PressureNLOpart}), supplemented by vacuum contributions 
Eqs(\ref{eq:PLOrgopt}),(\ref{eq:PNLOrgopt}), these expressions are modified by Eq.(\ref{delta}), expanded to LO (NLO) $\delta^0$ ($\delta^1$), then finally taking $\delta\to 1$ to formally recover the massless limit. 
At LO, the RG equation being already used to fix Eq.(\ref{acrit}), 
one uses Eq.(\ref{opt}) to determine $\overline{m}_{\rm LO}$, and
RG invariance is exact at this one-loop order. For completeness, we give the LO mass gap 
obtained in \cite{rgopt_cold}:
\begin{equation}\label{eq:MassLO}
\overline{m}_{\rm LO}^2=\mu^2 \left(\frac{\sqrt{1+4c\left(\overline{m}_{\rm LO},\mu,g^2\right)}-1}{2c\left(\overline{m}_{\rm LO},\mu,g^2\right)}\right),
\end{equation}
with
\begin{equation}\label{eq:C_LO_Lmu_pf_def}
c\left(\overline{m}_{\rm LO},\mu,g\right)=\left(\frac{1}{2b_0 g^2}-\frac{1}{2}+L_{\mu}\right)^2 \  \ \ , \ \ \ L_{\mu}=\ln \left(\frac{\mu+p_F}{\Lambda}\right) .
\end{equation}
Applying next the same procedure (\ref{delta}) to the NLO pressure, 
Eqs.(\ref{eq:PLOrgopt}),(\ref{eq:PNLOrgopt}), Eqs.(\ref{opt}) or (\ref{RGred}) gave no real $\overline m$ exact solution\cite{rgopt_cold} on the full relevant $\mu$-value range. Invoking RG-invariance of the pressure under renormalization scheme change (RSC) up to perturbatively higher orders, one may slightly change the scheme to a one close to $\overline{\rm MS}$, such that a comparison remains perturbatively consistent, but 
modifying the RG equation coefficients to possibly recover real $\overline m$ 
solutions\cite{RGOPTals}. The NLO RSC is performed
on the arbitrary mass $m$ according to
\begin{equation}\label{eq:B2}
m \rightarrow m \left(1+B_2\ g^4 \right),
\end{equation} 
to be applied prior to the variational modifications from Eq.(\ref{delta}), so that the RSC is defined in a standard perturbative manner. 
Notice that the higher order RG coefficients are also modified under the RSC 
(see Appendix A). 
Moreover, a definite prescription is needed to uniquely fix the RSC parameter $B_2$, 
arbitrary at this stage. Following\cite{RGOPTals,rgopt_cold}, 
the real solution closest to $\msbar$ is obtained
when the two independent OPT and RG equations, Eqs.(\ref{opt}),(\ref{RGred}), first intersect, 
respectively viewed as functions $f_{\rm OPT}(m,g^2)\equiv 0$ and $f_{\rm RG}(m,g^2)\equiv 0$ 
(i.e. when their respective tangent vectors are collinear). 
The latter prescription easily translates into a vanishing determinant condition:
\begin{equation}\label{eq:FRSC}
 \frac{\partial f_{\rm RG}}{\partial g^2}\frac{\partial f_{\rm OPT}}{\partial m}-
 \frac{\partial f_{\rm RG}}{\partial m}\frac{\partial f_{\rm OPT}}{\partial g^2} \equiv 0 .
\end{equation}
Since the latter equation depends nontrivially on $B_2$,
one solves it in conjunction with either Eq.(\ref{opt}) or Eq.(\ref{RGred}) to obtain $(\overline B_2, \overline m)$ solutions.
Following the previous prescription, the NLO RGOPT pressure was obtained in \cite{rgopt_cold},
that we will not reproduce here. 
We simply remark that due to embedded RG invariance properties of $\overline m(g,\mu)$, the resulting pressure exhibits 
a more moderate sensitivity to residual renormalization scale variations\cite{rgopt_cold},
although the improvement is not as drastic as in the analogous construction at finite temperature
and zero density\cite{rgopt_hot}.

\subsection{NNLO RGOPT}\label{rgoptnnlo}

\begin{figure}[h!]
	\centering
    \epsfig{file=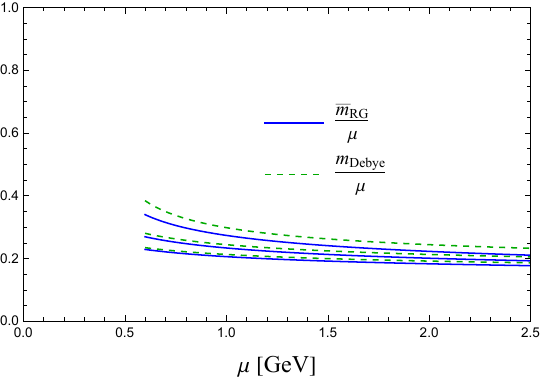,width=0.58\linewidth}
    \caption{Residual scale dependence of the RG-determined mass
    $\overline{m}_{\rm RG}(g(\Lambda) )$ for $\Lambda= 2\mu X$ with $X = \frac{1}{2},1,2$ (from top to bottom curves respectively), compared with the LO Debye mass.\label{Fig:mscale}}
\end{figure}

\begin{figure}[h!]
	\centering
\epsfig{file=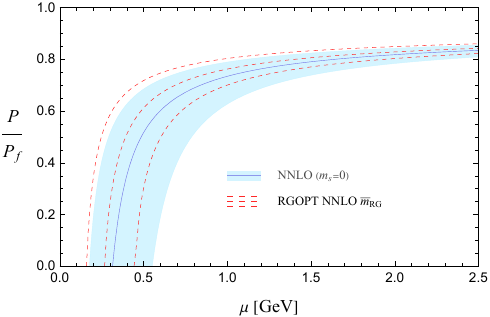,width=0.7\linewidth}
    \caption{The RGOPT pressure $P_{\rm RGOPT}^{N_f=3^*}(\overline m_{RG})$, Eq.(\ref{PRGOPT_NNLO}), compared to the NNLO pQCD pressure (massless and massive quarks as indicated) for 
    three different scales $\Lambda= 2\mu X$ with $X \in \[\frac{1}{2},1,2 \]$ (from bottom to top curves)  \label{Fig:PressureNNLO}}
\end{figure}

Next at NNLO, we include the RGOPT modifications of the relevant contributions in Fig. \ref{Fig:GraphNNLO} as well as the plasmon contribution in Fig. \ref{Fig:VV_VM_MM},
where the ring diagram, represented in Fig. \ref{Fig:Ring}, 
is given by Eq.(\ref{eq:Ring3star}).
To recap, the perturbatively NNLO RG-invariant massive quark pressure $P_{RGI}^{v+m}$ 
as obtained in Eq.(\ref{sub}),(\ref{eq:s_i Quark}) entails the contributions 
from Eqs.(\ref{Ph+v}), with the NNLO contributions (\ref{eq:PNNLOrgopt}),(\ref{eq:P32GIVM}).
Then one applies on $P_{RGI}^{v+m}$ the modifications from Eqs.(\ref{delta}),(\ref{acrit}) in the quark sector, 
these previously defined steps being formally summarized as:
\be\label{PRGOPT_NNLO}
P_{RGI}^{v+m}(m \to m (1-\delta)^{\frac{\gamma_0}{2b_0}},g^2 \to \delta g^2) |_{\delta^2,\delta\to 1}
\equiv P_{\rm RGOPT}^{v+m}(m, g) \equiv P_{\rm RGOPT}^{N_f=3^*}(m,g) .
\ee
The explicit expression resulting from Eq.(\ref{PRGOPT_NNLO}) is not particularly 
illuminating so we do not display it here. 
Upon next using either the OPT Eq.(\ref{opt}) or the RG Eq.(\ref{RGred}) to determine $\overline{m}$, we found no real solutions in the full relevant $\mu$ range. 
Thus, similarly to the NLO case, we perform a perturbative RSC in order to obtain 
real and continuous solutions in the full $\mu$ range, 
considering the same RSC Eq.(\ref{eq:B2}) as was performed at NLO.
After the modified
expansion from Eq.(\ref{delta}), this induces contributions to the pressure that remain linear in $B_2$,
therefore easier to handle analytically.
Similarly to NLO, we can recover a real solution 
by requiring Eq.(\ref{eq:FRSC}) to be satisfied. As a technical remark, while Eqs.(\ref{opt}),(\ref{RGred}),(\ref{eq:FRSC}) all have a very involved 
nonlinear $m$-dependence at NNLO, note that the RSC parameter $B_2$ may be first trivially obtained from 
the relevant RG Eq.(\ref{RGred}), linear in $B_2$. Inserting the resulting solution 
$B_2(m,g,\cdots)$ into Eq.(\ref{eq:FRSC}) provides a single equation for $m$, once fixing the other 
relevant parameters $g(\Lambda),\mu, \Lambda$.  
In this way the whole procedure to determine $\overline m(g,\mu,\Lambda)$ is analytically straightforward, 
although somewhat involved numerically. \\ 
While our prescription defines $\overline m_{RG}(g,\mu)$ from an RG-optimized pressure, thus unrelated to the standard Debye mass,
defined from the pole of the quark propagator, we expect a consistent solution to behave at
small $g^2$ as a screening mass in the medium, $\overline m^2 \sim {\cal O} (g^2 \mu^2)$, upon perturbative re-expansion.
Accordingly, it is sensible to compare at least qualitatively our solutions with the $T=0$ 
perturbative Debye screening quark mass:
\begin{equation}
 m^2_{\rm D} = \frac{g^2\, C_F}{8\pi^2} \mu^2.
 \label{mdebye}
\end{equation}
It turns out that the $\overline m_{RG}(g,\mu)$ obtained from Eqs.(\ref{RGred}) and (\ref{eq:FRSC}) is
remarkably close to $m_D$ (at least for moderate and high $\mu$ values in the perturbative range), 
as illustrated in Fig.~\ref{Fig:mscale}. In contrast,  
using alternatively the OPT Eq.(\ref{opt}) with Eq.(\ref{eq:FRSC}) gives a NNLO solution more than an order of magnitude smaller 
than $m_D$ for any $\mu$ values, which accordingly we do not consider a physically acceptable solution\footnote{Note also that 
using instead of Eq.(\ref{eq:B2}) an highest order RSC at NNLO,
$m\rightarrow m \left(1+B_3\ g^6 \right)$,
only gives unphysical real solutions $\overline m = {\cal O}(\mu)$ instead of $\overline m = {\cal O}(g \mu)$.}.
The complete NNLO RGOPT implementation is summarized for convenience in Table \ref{tab-algo}.
Concerning the RSC parameter $\overline B_2(\mu)$ determined together with $\overline m_{RG}(\mu)$, for a given scale $\Lambda$ we obtain $B_2$ values almost constant for a large $\mu$ range, 
e.g. $\overline B_2 \sim -0.025$ for $\Lambda = 2\mu$. 
These relatively large values can be traced to 
the sizable (negative) departure from the massless quark NNLO pressure 
for a typical dressed mass $\simeq m_{\rm D}$, 
originating dominantly from $G_1(m/\mu), G_2(m/\mu)$ entering the 
NNLO massive contributions Eq.(\ref{eq:PressureM3}) (see Fig. \ref{Fig:Gi_Comparison}),
requiring a relatively sizable $|B_2|$ to recover real $\overline m_{RG}(\mu)$ solutions.
However, importantly 
the fact that $\overline m_{RG}^2 \sim {\cal O} (g^2\,\mu^2) $ for moderate $g$ guarantees
that the modifications induced in the pressure from the RSC, 
$\propto \overline B_2\, g^2 \overline m^4, \overline B_2\, g^4 \mu^2 \overline m^2$ or higher orders, remain perturbatively 
formally of higher order 
${\cal O}(g^6)$.\\
In Fig. \ref{Fig:mscale}  we also illustrate the residual scale dependence of $\overline m_{RG}$ as compared to the perturbative Debye mass, Eq.(\ref{mdebye}). The scale dependence is  moderately reduced, but once inserting 
$\overline m_{RG}$ in the RGOPT pressure, 
the latter exhibits a further reduced scale dependence,
compared to the NNLO weak-coupling expansion pressure, as observed in Fig. \ref{Fig:PressureNNLO}. 
The NNLO pressure also shows a better scale dependence with respect to NLO RGOPT\cite{rgopt_cold}.
Overall, the reduction of scale dependence from RG resummation we obtain for cold dense matter is, however, not as efficient as its counterpart for hot QCD\cite{rgopt_hot}. This may be explained by the already
improved scale dependence and convergence of the standard weak-coupling expansion for cold quark matter, as compared to its counterpart for hot QCD. \\
Since the RG dressed mass is very close to the Debye mass $m_{\rm D}$
in Eq.(\ref{mdebye}),  
one may perhaps consider using the latter as a much simpler alternative mass prescription (as it is indeed done 
at finite temperature and densities for NNLO HTLpt\cite{HTLptDense3L}). However, in this case the resulting pressure does not show any improved scale dependence with respect to the standard  NNLO pressure. Likewise, perturbatively reexpanding\cite{rgopt_cold} 
the $\overline m_{RG}$ solution would result in a degradation of the scale dependence, compared to
the exact $\overline m_{RG}$.
This illustrates that the RG-dressed mass is quite crucial to gain scale independence, due to the embedded higher RG order dependence. \\

On top of the previous RG improved NNLO results, one could wish to include a recently derived all order resummation of the HTL {\em soft} logarithmic dependence in the pure glue sector\cite{Fernandez:2021jfr}. More precisely, the ring contribution discussed earlier is only the leading order of
a sequence of non-analytic contributions in $g^2$ to the pressure. Going to higher orders, higher powers of $\ln g$ will appear due to IR divergences resummations\cite{Gorda:2021znl}, and those contributions arise with a specific pattern, that happens to be dictated by the RG in an adequately defined effective field theory (EFT) set-up\cite{Fernandez:2021jfr}. This EFT is built on the hard thermal loops generated by the gluons\cite{HTLbasic}, hence the name HTL EFT. At the moment, only the leading logarithms  ($g^4 \ln g,\ g^6 \ln^2 g,\ g^8\ln^3 g,\dots$) are completely resummed in this approach. However, it appears presently difficult to combine the gluon sector resummation results
in \cite{Fernandez:2021jfr} with the RGOPT approach performed here, 
as it would require to apply a similar variational procedure rather in the HTL EFT, with an
effective gluon mass relevant for dense HTL. 
But missing higher order contributions prevents us from pursuing such considerations here, 
which are beyond our present scope and left for future work.

\section{Incorporating the Strange quark mass in RGOPT: $N_f=2^*+1^*$}
\subsection{Non-diagonal massive contributions at NNLO}
So far, we discussed fully symmetric $N_f=3$ quark matter with RG-dressed masses 
$m_1= m_2= m_3\equiv \overline m_{RG}$ and chemical potentials 
$\mu_1=\mu_2=\mu_3\equiv \mu$. 
Reintroducing different chemical potentials is straightforward within the NNLO 
contribution $\mathcal{M}_3^{N_f=3^*}$, Eq.(\ref{eq:P32GIVM}), but more involved in the ring contributions in Fig.~\ref{Fig:Ring}. However, when accounting for beta-equilibrium 
and charge neutrality, since the electron chemical potential $\mu_e \ll \mu \equiv \mu_d = \mu_s $, 
it leads to very small differences among the three respective quark $\mu_i$ values. Moreover, the ring contribution being itself numerically subdominant for a large $\mu$ range with respect to other NNLO contributions, accounting for these differences would lead to completely negligible effects in the total pressure. Indeed, we anticipate that the strange quark mass effects, on top of
the RG-dressed mass, are numerically quite irrelevant within the ring contributions, 
while the latter effects are typically an order of magnitude larger than unequal chemical potential contributions. After a lengthy numerical evaluation we found these modifications to be too small to be relevant at all. Therefore, as far as the ring contributions are concerned, 
for computational efficiency 
we stick to the $N_f=3^*$ massive ring results given earlier in \Eq{\ref{eq:Ring3star}}.

In contrast, concerning the other largely dominant NNLO contributions to the pressure, 
Eqs.(\ref{eq:P32GIVM}),(\ref{eq:PNNLOrgopt}), our goal now is to treat a more general case $(m_1= m_2=\overline m_{RG},\, m_3=\overline m_{RG}+\mcm)$, where $\mcm$ is the genuine strange quark (running) mass. 
We anticipate that the latter is the relevant mass pattern within NNLO contributions,
resulting from the appropriate generalization of Eq.(\ref{delta}) when incorporating $m_s \ne 0$,
as will be clear below.
Starting at $\mathcal{O}(\alpha_s^2)$, there are new ``mixed'' (non-diagonal) contributions from having nondegenerate quark masses. This is easily observed from the plasmon displayed in Fig. \ref{Fig:VV_VM_MM} where two loops of independent quarks contribute.
These effects from unequal masses are not a priori negligible, in particular for relatively
low $\mu$ values where the NNLO contributions $\sim g^4(\Lambda\sim \mu)$ are enhanced, and
at the same time the strange quark running mass $m_s(\Lambda\sim \mu)$ and
RG-dressed mass $\overline m_{RG}(g,\mu)$ values may become roughly of similar order.
The first contribution is purely a vacuum (VV) contribution, which has to be incorporated in \Eq{\ref{eq:PNNLOrgopt}}. The second graph is incorporated in an appropriately modified $\mathcal{M}_3^{N_f=3^*}$ of Eq.(\ref{eq:P32GIVM}), while the last one is part of the ring resummation. For a careful derivation of the modified vacuum contribution, see Appendix C. Collecting the result from \Eq{\ref{eq:PvacNNLOmix}}, we find the replacement 
\be
\begin{aligned}\label{VVnd}
  N_f P^{v,N_f=3^*}_{\rm NNLO}\to P^{v,N_f=2^*+1^*}_{\rm NNLO}&= 
  \left(\frac{g^4}{\pi^2\,(4\pi)^4} \right) \bigg( (N_f-1)m^4 
  \( -156.833 + 253.775 L_m - 243 L_m^2 + 100 L_m^3 \)  \\
  & \left. +2(N_f-1)m^2 m_3^2 \(-4.75802 - 6 \(L_m + L_{m_3}\) \)  \right. \\
  & +m_3^4 \(-152.075 + 265.775 L_{m_3} - 243 L_{m_3}^2 + 100 L_{m_3}^3\) \bigg) , 
  \end{aligned}
\ee
accounting for the $N_f^2$ degrees of freedom, with $L_m=\ln(m/\Lambda)$ and $L_{m_3}=\ln(m_3/\Lambda)$. Accordingly, the zero point energy 
terms in \Eq{\ref{sub}} must be modified similarly such as to preserve perturbative RG invariance, according to
\be
\begin{aligned}\label{submix}
-N_f \frac{m^4}{g^2} \sum_{k=0}^{2}s_k\,\(g^2\)^k \to P_{\rm sub}^{N_f=2^*+1^*}\equiv &-(N_f-1)\frac{m^4}{g^2}\(s_0+s_1\,g^2\)-\frac{m_3^4}{g^2}\(s_0+s_1\,g^2\)\\
&-(N_f-1)s_{2,1}m^4\,g^2-s_{2,3} \, m_3^4\,g^2-2(N_f-1)s_2^{nd} \, m^2 m_3^2\, g^2.
\end{aligned}
\ee
(See appendix A for the explicit expressions of the $s_{i,j}$ coefficients).

The derivation of the ``VM'' graph with three different masses is more involved:
details about the modifications implied for this new $\mathcal{M}_3^{N_f=2^*+1^*}$ are given in Appendix C.2. To keep track of the origin of the mass coming either from the vacuum or the matter loop, we rename the latter $m\to m_i$ according to $\mathcal{M}_3^{N_f=3^*}(m)\to \mathcal{M}_{3}^{N_f=2^*+1^*}(m_i)$. In summary, to switch from the fully symmetric massive quarks ($N_f=3^*$) case to the more general $N_f=1^*+1^*+1^*$ case with different masses (from which $N_f=2^*+1^*$ is a specific case) amounts 
to apply the following modifications of the last terms in \Eq{\ref{eq:PressureM3}}\footnote{We recall
that $G_4(x)$ enters the $N_f$ coefficient when going from $N_f=2+1^*$ to $N_f=3^*$, see Eq.(\ref{M1c2cmod}).}:
\be\label{eq:ModifM3_2masses}
\begin{split}
N_f\(\frac{2}{3}\ln\frac{\mh}{2}+\frac{11}{9}+G_3(\mh)+G_4(\mh)\) \to & N_f\(\frac{2}{3}\ln\frac{\hat{m}_i}{2}+\frac{11}{9}+G_3^{\rm bis}(\mh_i)\)+\mh_i^2 z(m_i)\sum_{j}^{N_f}\Bigg\{{\rm Li}_2(\nui)\(1+\nui^{-2}\)\\
&-\Phi\(\nui,2,\frac{3}{2}\)\(1+\nui\)-\nui^{-2}+\mathcal{F}(\nui)\Bigg\}-\sum_{j}^{N_f}\frac{4}{3}I_{12}^{\rm bis}(\mh_i,\nui),
\end{split}
\ee
where  $\nui=(m_i/m_j)^2$, $\mathcal{F}(\nui)$ is defined in \Eq{\ref{eq:VM1cPlusFnui}}, $\Phi(a,b,c)$ is the Lerch Zeta function, ${\rm Li}_{2}$ is the Polylogarithm function and we conveniently provide $G_3^{\rm bis}$ as a rather accurate fitting function, quite similarly to those in Eq.(\ref{eq:GiFunc}):
\be
\hspace{-0.8cm}
G_3^{\rm bis}(\mh)= 32\pi^4\mh^2 \Big( -0.000244-0.003777 \uh +0.000319\uh^2+0.001263\uh^3  
 +0.000322 \ln\mh +0.000572\ln^2\mh+0.003743\mh^2\ln\(\frac{1+\uh}{\mh} \) \Big) .
\ee
In addition, the definition and a more convenient two-dimensional fit of the integral $I_{12}^{\rm bis}$ is given in Eq. (\ref{eqC13}). To recover the massless limit, one must first take the limit $\nui\to 1\, \forall\: i,\, j$, where all masses
become identical, and only then taking the limit $m\to 0$. (Upon taking the former limits, we correctly reproduce the left-hand side of \Eq{\ref{eq:ModifM3_2masses}} as a crosscheck).\\
Finally to recap, the complete $N_f=2^*+1^*$ QCD pressure reads
\be
\begin{aligned}\label{Prgopt_ms}
\mathcal{P}_{\rm QCD, NNLO}^{N_f=2^*+1^*}(m,m_s,\mu)=&(N_f-1)\(P_{\rm LO}(m,\mu)+P_{\rm NLO}(m,\mu)+P_{\rm 2GI,VM}^{N_f=2^*+1^*}(m,\mu)\)+P_{\rm LO}(m_3,\mu)+P_{\rm NLO}(m_3,\mu)\\
&+P_{\rm 2GI,VM}^{N_f=2^*+1^*}(m_3,\mu)-\Omega_{\rm Ring}^{N_f=2^*+1^*}+P_{\rm NNLO}^{v,N_f=2^*+1^*}+P_{\rm sub}^{N_f=2^*+1^*},
\end{aligned}
\ee
where, as above explained, we made the legitimate approximation $\Omega_{\rm Ring}^{N_f=2^*+1^*}\approx\Omega_{\rm Ring}^{N_f=3^*}$.
\subsection{NNLO RGOPT pressure for $N_f=2^*+1^*$}
From here, we apply the RGOPT procedure as detailed in section III, Eqs.(\ref{delta})-(\ref{RGred}),
with the modification that Eq.(\ref{delta}) is now generalized to
\be\label{deltams}
\begin{aligned}
&m_{i}\to m \left(1-\delta\right)^{\fr{\gamma_0}{2b_0}}, \ i=1,2\ \\
&m_3\to m\(1-\delta\)^{\fr{\gamma_0}{2b_0}} +m_s \\
&\; g^2\to \delta g^2,
\end{aligned}
\ee
accordingly importantly the genuine physical mass $m_s$ remains unaffected by the $\delta$-expansion. This leads to a corresponding generalization of Eq.(\ref{PRGOPT_NNLO}) for $m_s\ne 0$.
For completeness, the different steps are also summarized in Table \ref{tab-algo}.
\begin{table}[h!]
\renewcommand{\arraystretch}{1.1}
\begin{center}
\begin{tabular}{|c|c|c|c|}
\hline
Pressure model   &   Methods   &  Contributions &  Relevant Eqs. \\
\hline\hline
  & NNLO  & $P^m$ =NNLO ($m\ne 0$), $N_f\times m$  &
 (18)...(21),(22),\sout{(11)} \\
\cline{3-4}
 $N_f=3^*$ & weak expansion &  + NNLO $m \ne 0$ vacuum terms &
                     (23)...(25),(26) \\
                     \cline{2-4}
  (medium + vacuum)  & RG invariance  & + RGI-restoring contributions & (33),(A9),(A10) \\
\cline{3-4}
$P^m +P^v$ &   & = NNLO $P^{v+m}_{\rm RGI}(m)$ & (28) \\
\hline
          & NNLO RSC(m) & NNLO $P^{v+m}_{\rm RGI}(m,B_2)$ & (40) applied to (28)\\
\cline{2-4}
$N_f=3^*$ RGOPT  & RGI $\delta$-expansion & modified $P^{v+m}_{\rm RGI}(m,B_2)$  &   (34),(35) @NNLO$(\delta)$ \\
\cline{2-4}
             & NNLO RG Eq.   & RG-dressed $\overline m(g,\mu)$  &  (37),(41) \\
\cline{3-4}
        &   &
        $\rightarrow$ NNLO $P^{v+m}_{\rm RGI}(\overline m(g,\mu))$
        & (42)($m\to \overline m(g,\mu)$)\\
\hline
\hline
$N_f=2^* +1^*$  & & + NNLO $m_i \ne m_j$ terms & (44)...(46),(47),(48) \\
\hline\hline
              & NNLO RSC(m, $m_s$) & NNLO $P^{v+m}_{\rm RGI}(m,B_2,m_s)$ & (40) applied to (48)\\
\cline{2-4}
$N_f=2^* +1^*$ RGOPT &  RGI $\delta$-expansion  & modified $P^{v+m}_{\rm RGI}(m,m_s,B_2)$ & (49),(35) @NNLO$(\delta)$  \\
\cline{2-4}
& NNLO RG Eq. & RG-dressed $\overline m(g,\mu)$, $m_s \ne 0$    &   (29),(41)\\
\cline{3-4}
        &      & $\rightarrow$ NNLO $P^{v+m}_{\rm RGI}(\overline m(g,\mu),m_s)$
        & (51)\\
\hline
\end{tabular}
\end{center}
\caption{Summary of steps and patches for
$N_f=3^*$ or $N_f=2^* +1^*$ NNLO RGOPT for the cold dense QCD pressure. Within those different
``models'' the prescriptions flow from top to bottom, with relevant expressions in the rightmost column.
Numbers in parenthesis refer to equations in the main text, with $+$
or $...$ signs indicating that equations are summed, while a comma indicates extra definitions from the
subsequent equations.
Abbreviations: 
\sout{(11)}: removed; RGI: RG invariant; RSC: renormalization scheme change.}
\label{tab-algo}
\end{table}
%
Upon recovering $\delta\to 1$ after the appropriate $\delta$-expansion being performed up to NNLO
($\delta^2$), notice that 
the originally NNLO $m$-dependent contributions, being already ${\cal O}(g^4)$,
are simply obtained from $m\to m+m_s\equiv m_3$ concerning the strange quark contributions, 
due to the last term in Eq.(\ref{deltams}). Accordingly, this brings NNLO contributions
of the form derived above in Eqs.(\ref{VVnd}), (\ref{submix}), (\ref{eq:ModifM3_2masses}) as anticipated. 
An additional important
modification as compared to $m_s =0$ is that the RG operator has to be consistently extended 
by introducing the strange quark anomalous mass dimension within \Eq{\ref{RGred}}, 
\be\label{eq:RGredms}
\Lambda\frac{d}{d\Lambda}=\Lambda\frac{\partial}{\partial\Lambda}+\beta(g^2)\frac{\partial}{\partial g^2}-\mcm \gamma_{\mcm} \frac{\partial}{\partial\mcm}.
\ee
Note indeed that if restricting Eq.(\ref{Prgopt_ms}) to LO, Eq.(\ref{deltams}) breaks explicitly, 
by ${\cal O}(m_s/m)$ terms, the exact RG-invariance of the LO pressure
previously satisfied for arbitrary $m$, resulting from the critical exponent in Eq.(\ref{acrit}). 
However, similarly to our NLO and NNLO prescription for $m_s=0$, the massive LO RG invariance can be
simply recovered upon reimposing Eq.(\ref{eq:RGredms}): 
it gives an already nontrivial LO screening mass solution, behaving for sufficiently small $m_s$ as 
$\overline m_{RG,LO}\sim {\cal O} (g \mu) +{\cal O}(m_s)$.
Next proceeding at NNLO, applying the complete RGOPT prescription with $m_s \ne 0$ 
is slightly more involved but conceptually very similar to the procedure described above in Sec.\ref{rgoptnnlo}, 
so that we basically use now the massive RG Eq.(\ref{eq:RGredms}), after having performed the RSC according to Eqs.(\ref{eq:B2}),(\ref{eq:FRSC}), the latter being required in order to recover real 
$\overline m_{RG}$ solutions. 
Due to nonlinear $m$, $m_s$ dependencies, the solutions $\overline m_{RG}(m_s)$ 
are not related in a simple manner to $\overline m_{RG}(0)$ obtained in Fig. \ref{Fig:mscale}, 
but one always get $\overline m_{RG}(m_s) < \overline m_{RG}(0)$, which can be understood since $m_s \ne 0$ gives additional positive contributions 
to the pressure. As expected, this effect is relatively small 
for very perturbative $\mu, \Lambda$ values,
while it becomes more important for very low $\mu$ and $\Lambda=\mu$, i.e.
large $\alpha_s(\Lambda)$ values.\\

\begin{figure}[h!]
	\centering
\epsfig{file=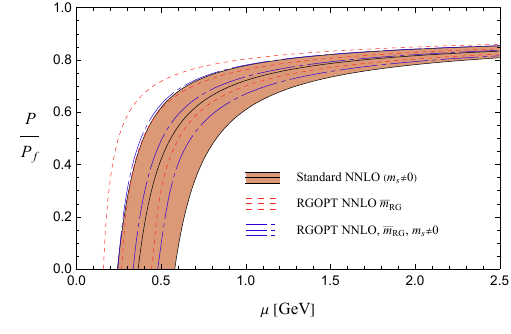,width=0.7\linewidth}
    \caption{The RGOPT $N_f=2^*+1^*$ pressure compared to the NNLO pQCD  $N_f=2+1^*$ pressure  for 
    $\Lambda= 2\mu X $ with $X \in \[\frac{1}{2},1,2\] $ (from bottom to top curves).
     \label{Fig:PressureNNLOms}}
\end{figure}

Inserting the obtained $\overline m_{u,d} \equiv \overline m_{\rm RG},\
\overline m_3\equiv \overline m_{\rm RG}(m_s)+m_s$ masses into Eq.(\ref{Prgopt_ms}) 
gives our final result for the $N_f=2^*+1^*$ pressure, accounting also for the 
running $m_s(\Lambda)$ from Eq.(\ref{msrun}).
The resulting pressure with its remnant scale dependence is displayed in Fig. \ref{Fig:PressureNNLOms}, compared with the standard NNLO pQCD pressure ($N_f=2+1^*$), Eq.(\ref{eq:P2+1NNLO}), and with the RGOPT pressure for symmetric quark matter ($N_f=3^*$), 
$P_{RGOPT}^{v+m}(\overline m_{RG})$ in Eq.(\ref{PRGOPT_NNLO}).  
Importantly, we observe that our result lies within the uncertainty range of the standard NNLO 
QCD pressure but with a significantly reduced scale dependence with respect to the latter.
Remark that the {\em relative} difference between respectively the resummed RGOPT pressures
for massless quark and for $m_s \ne 0$ appears rather 
important for low $\mu$ values, as could be expected since $m_s(\Lambda)/\mu$ is not that small. For instance, for the central scale $\Lambda=2\mu$ and $\mu =0.5$ GeV, where $\alpha_s\simeq 0.4$, 
the RGOPT pressure for $m_s \ne 0$ is reduced by $\sim 20\%$ with respect to the massless pressure.
This effect, however, is roughly comparable to the corresponding 
$m_s \ne 0$ effect for the standard NNLO pressure, comparing the latter 
in Fig. \ref{Fig:RelativeErrorPQCDnewGi} (left).\\
We remark also that the threshold of the Heaviside $\theta(\mu-(\overline{m}+m_s))$ function is never reached, down to the lowest $\mu$ values here considered, therefore, the strange quark always populates the quark matter medium. The RGOPT pressure for the central scale ($\Lambda=2\mu$) reaches zero value for a somewhat smaller critical $\mu_c(\approx 0.338$ GeV) than the pQCD one ($\mu_c\approx 0.364$ GeV). This, in addition with the reduced scale dependence, is expected to 
have interesting consequences for the EoS relevant to compact stars. Such considerations, however, are beyond the scope of this work.
\subsubsection{Pocket formula for the RGOPT $N_f=2^*+1^*$ pressure}

Finally, the procedure to reproduce our final results for $N_f=2^*+1^*$ being rather involved, we provide a simpler pocket formula as a good fit to the numerical RGOPT pressure result in Fig. \ref{Fig:PressureNNLOms}, inspired from a similar construction in \cite{prevN2LOfit}:
\ba
&& \frac{\mathcal{P}_{\rm RGOPT}^{N_f=2^*+1^*}(\mu,\Lambda=X\, \mu)}{\mathcal{P}_{f}(\mu,N_f=3)}=\(c_1 + c_2  X^{\nu_3}\) - \frac{ d_1   (3 \Tilde{\mu})^{\alpha_1}X^{\nu_1}}{(3 \Tilde{\mu} - d_2\, X^{-\nu_2}) },\ \ \  \Tilde{\mu}=\mu/{\rm GeV}, \nn
&& c_1=0.766035 , c_2=0.501495 , \alpha_1=0.996305, d_1=0.402405, d_2=0.974897, \nn  
&& \nu_1=0.410395,\ \nu_2=0.631054,\ \nu_3=0.366230.
\ea
\section{Summary and conclusion}
In this work we have applied the RGOPT resummation approach at NNLO to the cold quark matter pressure. 
As a preliminary basic ingredient of this approach, 
we have re-investigated the derivation of the NNLO massive cold quark matter pressure for two massless and one massive quark ($N_f=2+1^*$), originally  evaluated in \cite{Kurkela:2009gj}. While we have reproduced all analytical intermediate and final results from
the latter work, we obtain a mismatch in one of the numerical fitting function ($G_2$) 
for massive integrals which, however, has a small impact on the standard NNLO pQCD pressure result, due to the moderate $m_s(\Lambda)$ relevant values. 
Then, we have proceeded to derive the RGOPT resummation at NNLO,
first for the simpler degenerate $N_f=3^*$ case, and next for the more realistic $N_f=2^*+1^*$ case,
the latter incorporating effects from the genuine strange quark mass. 
This involves first some modifications required to incorporate a fully symmetric massive pressure ($N_f=3^*$), followed by the more general case of non-symmetric massive pressure ($N_f=1^*+1^*+1^*$). 
Next, reintroducing vacuum contributions play a crucial role in determining a perturbative RG invariant pressure,
as examined in Sec. III.B. Importantly also,
the modified perturbation from Eq.(\ref{delta}) is maintaining perturbative RG invariance, and
 the NNLO RGOPT implementation involves a sequence of  well-defined 
steps (summarized in Table \ref{tab-algo}) leading to a unique RG-dressed mass 
compatible with a screening behavior.
For both $N_f=3^*$ case or $N_f=2^*+1^*$ with $m_s\ne 0$, generically
our procedure embeds a higher order RG dependence, thus adding higher order $\Lambda$-dependent contributions
that can further reduce the renormalization scale dependence as generically expected from RG cancellations, so that
our results display a significantly reduced scale dependence with respect to the standard NNLO pressure \cite{Kurkela:2009gj}. The scale dependence reduction is nevertheless limited,
since higher order perturbative coefficients remain unknown anyway. Accordingly,
the approach has the same limitations and primary sources of uncertainties than standard perturbation. 
In principle, some possible variant of our prescriptions 
could potentially add theoretical uncertainties inherent to such variational approaches. 
Although we obviously cannot exclude the latter, we do not examine possible extra uncertainties that could affect our final results, since the NNLO RGOPT prescription 
is uniquely defined.
The latter RG resummation improvements appear, however, not as efficient 
as their counterpart for hot QCD\cite{rgopt_hot} or for a hot scalar theory\cite{rgopt_phi4_NNLO} at NNLO. This is likely explained by the already better scale dependence and convergence properties of the weak-coupling expansion for cold quark matter, as compared typically to hot QCD.
The RGOPT improvements should follow up for the EoS of cold quark matter in the same approximation, thereby potentially reducing the present pQCD uncertainties in the intermediate $\mu_B$ regime very
relevant for the properties of compact stars. 
\section{Acknowledgments}
We are very thankful to Aleksi Vuorinen for discussions, and for providing us useful intermediate numerical results related to \cite{Kurkela:2009gj} for more precise comparisons. L.F. has been supported in part by the Research Council of Finland Grants no. 353772 and 354533. L.F. is also grateful to the University of Montpellier where this work was started.

\appendix

 \section{}
 \subsection{Renormalization group functions and counterterms}
We give here relevant expressions for the coefficients of the RG functions $\beta(g^2)$, $\gamma_m(g^2)$ as well as the subtraction coefficients entering $S(m,g)$ in Eq.(\ref{sub}). 
Up to three-loop order (NNLO) one has
\begin{equation}
 \beta\left(g^2\equiv 4\pi\alpha_s \right)=-2g^4\left(b_0 +b_1g^2+b_2 g^4+ 
 \mathcal{O}\left(g^6\right) \right)\;,
 \end{equation}
 \begin{equation}
 \gamma_m\left(g^2\right)=g^2\left( \gamma_0+\gamma_1g^2 +\gamma_2 g^4 +
 \mathcal{O}\left(g^6\right)\right)\;,
\end{equation}
with the successive order coefficients, for the relevant QCD case with $C_A=N_c$, $C_F=4/3$, $N_c=3$, 
but keeping the number of quark flavors $N_f$ unspecified:
\begin{equation}\label{eq:BetaCoef}
\begin{aligned}
b_0= & h\left(\fr{11}{3}C_A-\frac{2}{3}N_f\right) \ \ ,\ \ 
b_1 =  h^2 \left(\fr{34}{3} C_A^2 - 2 C_F N_f - \fr{10}{3} C_A  N_f \right) \ \ , \ \ 
b_2 = h^3 \left(\frac{2857}{2}-\frac{5033}{18}N_f+\frac{325}{54}N_f^2\right)\\
b_3 = & h^4 \left(\frac{149753}{6}+3564 \zeta(3) - \Big( \frac{1078361}{162}+\frac{6508}{27}\zeta(3)\Big) N_f+\Big( \frac{50065}{162}+\frac{6472}{81}\zeta(3)\Big) N_f^2 +\frac{1093}{729} N_f ^3 \right)
\end{aligned}
\end{equation}
with $h=1/(4\pi)^2$,
\begin{equation}\label{eq:GammaCoef}
\begin{aligned}
\gamma_0 = & 2h (N_c C_F ), \ \ \ 
\gamma_1= 2h^2 \left(\fr{3}{2} C_F^2 + \fr{97}{6} C_F C_A - \fr{5}{3} C_F N_f\right), \ \ 
\gamma_2=  2 h^3\left(1249+ N_f \left(\frac{-2216}{27}-\frac{160}{3}\zeta (3)\right)-\frac{140}{81}N_f^2\right) \\
\gamma_3 = & 2 h^4 \Big(\frac{4603055}{162}+\frac{135680}{27} \zeta(3)-8800 \zeta(5)+ \Big(-\frac{91723}{27}-\frac{34192}{9}\zeta(3)+880 \zeta(4)+\frac{18400}{9}\zeta(5)\Big) N_f\\
& + \Big(\frac{5242}{243}+\frac{800}{9}\zeta(3)-\frac{160}{3} \zeta(4)\Big) N_f^2+ \Big(-\frac{332}{243}+\frac{64}{27}\zeta(3)\Big) N_f^3 \Big). \\
\end{aligned}
\end{equation}
Next, the coupling and mass counterterms are defined in our normalization conventions
and with $D=4-2\e$, as
\ba\label{ZgZm}
g_0^2 &\equiv& g^2 Z_g = g^2 \left(1 -  g^2 \fr{b_0}{\e} + g^4 (\fr{b_0^2}{\e^2} - 
\fr{b_1}{2 \e}) +{\cal O}(g^6)\right) \nn
m_0 &\equiv& m Z_g = m \left(1 - g^2\fr{\gamma_0}{2 \e} + 
 g^4 ( \fr{\gamma_0 (2 b_0 + \gamma_0)}{8 \e^2} - \fr{\gamma_1}{4 \e} )       
 +{\cal O}(g^6)\right), 
\ea
where $g_0$, $m_0$ are the bare coupling and mass.
Without going into full details, we simply remark that essentially taking the LO and NLO contributions in Eqs.(\ref{eq:PressureLOpart}),(\ref{eq:PressureNLOpart})
formally with $m\to m_0$, $g\to g_0$ and using Eq.(\ref{ZgZm}) cancels the explicit UV divergences
initially present in the 1-cut and 2-cut contributions (see Eqs.(\ref{lnZ2GI2c}), (\ref{eq:VM1cPlusFnui}), (\ref{OmVM2cbare})), resulting into the finite contributions Eqs.(\ref{1cut}),(\ref{2cut}).\\
The vacuum energy counterterm $Z_0$ is defined as\cite{vacEn2,vacEn5l} 
\be
{\cal E}_0^B \equiv \Lambda^{-2\e} \left( {\cal E}_0(g^2)-m^4\,Z_0(g^2) \right)
\ee
where ${\cal E}_0^B, {\cal E}_0$  are respectively the bare and renormalized vacuum energies and $Z_0$ the dimensionless counterterm (that cancels divergences originally present only in the vacuum
contributions, that are not canceled after coupling and mass renormalization).
From this follows the vacuum energy anomalous dimension, that reads in our normalization:
\be
\hat\Gamma^0(g^2) = 
(-4\gamma_m(g^2) -2\e) Z_0(g^2) +\overline\beta(g^2) \frac{\partial}{\partial g^2} Z_0(g^2)
\equiv \sum_{k\ge 0}  \Gamma^0_k g^{2k} 
\label{vacen}
\ee
with $\overline \beta(g^2)\equiv -2\e g^2+\beta(g^2)$ as it is appropriate for bare quantities. 
Explicitly, the vacuum energy anomalous dimension coefficients are (translated into our normalization from \cite{vacEn2,vacEn5l})  
\begin{equation}\label{eq:vacEncoeff}
\begin{aligned}
&\Gamma^0_0= -2h N_c, \ \ 
\Gamma^0_1=-2h\frac{N_c}{(4\pi^2)} C_F, \ \ 
\Gamma^0_2=-2h\frac{N_c}{(4 \pi^2)^2} 
\left(\frac{457}{72}-\frac{29}{12}N_h-\frac{2}{3}\zeta(3)-\frac{5}{12}N_l \right),\\
& \Gamma^0_3= -2h\frac{N_c}{(4\pi^2)^3}
\left( 33.6625-32.5586 N_h+0.18139 N_h^2+0.214632 N_h\, N_l-4.96507 N_l+0.0332417 N_l^2 \right),
\end{aligned}
\end{equation}
where generically $N_h, N_l$ are respectively the number of massive and massless flavor of quarks.

Note that the NLO in Eq.(\ref{ZgZm}) is sufficient for the mass renormalization counterterm within the medium 
contributions, 
while higher order RG coefficients enter the RG-restoring coefficient $s_k$ in Eq.(\ref{sub}), 
those being only relevant for the pure vacuum contributions.  
These $s_k$ 
are determined from Eq.(\ref{sub}) or equivalently alternatively from Eq.(\ref{eq:vacEn})
with Eq.(\ref{eq:vacEncoeff}),
which gives explicitly\cite{KNcond}\footnote{The present normalization is different from the one used in \cite{KNcond}, where the $s_k$ coefficients were defined for the vacuum quark condensate. 
The latter is related to the vacuum contribution to the pressure, $P(m)$, as
$\langle q\bar{q}\rangle \equiv -\partial_m P(m)$. Note also that Eq.(\ref{eq:s_i Quark})
holds for the pressure {\em per} flavor: accordingly the $s_k$ in Eq.(\ref{eq:s_i Quark})
receive an extra overall factor $N_f$ for $N_f$ degenerate massive quarks.}
\begin{equation}\label{eq:s_i Quark}
\begin{aligned}
s_0 = &\frac{\Gamma^0_0}{2\big(b_0-2 \gamma_0\big)}=\frac{3}{7} \ \ , \ \ 
s_1 = \frac{1}{2\gamma_0}\left((b_1-2 \gamma_1) s_0 -\frac{\Gamma^0_1}{2}\right)=-\frac{53}{224\pi^2}\\
s_2(N_h,N_l) = & \frac{1}{b_0+2\gamma_0}\left((b_2-2\gamma_2)s_0-
2\gamma_1 s_1 -\frac{\Gamma^0_2}{2}\right) \underset{N_h=3,N_l=0}{=} -0.00040082 \\ 
s_3 (N_h,N_l) = & \frac{1}{2(b_0+\gamma_0)}\left((b_3-2\gamma_3)s_0-2\gamma_2 s_1-
(b_1+2\gamma_1) s_2 -\frac{\Gamma^0_3}{2}\right) \underset{N_h=3,N_l=0}{=} -0.00008304 .
\end{aligned}
\end{equation}
Next, for the unequal quark mass case, dubbed as $N_f=2^*+1^*$ in the main text, 
the appropriate RG-restoring subtraction
terms, as defined in Eq.(\ref{submix}), are determined as
\be
\begin{aligned}
s_{2,1}=s_2(N_h=2,N_l=1)=-0.00024683, \ \ \ s_{2,3}=s_2(N_h=1,N_l=2)=-0.00009284,\ \ s_2^{nd}=\frac{9}{8\pi^4(2N_f-81)}
\end{aligned}
\ee
\subsection{Renormalization scheme change}
For the relevant RSC, Eq.(\ref{eq:B2}), one should account that it also affects RG coefficients\cite{Collins}. Accordingly, this induces
the following modifications\cite{RGOPTals}\footnote{Notice that Eq.(\ref{eq:B2}) is
the reciprocal RSC as compared to the one defined in \cite{RGOPTals}, thus some opposite signs occur in these relations.}:
\begin{equation}
\begin{aligned}
    \gamma_2&\to \gamma_2^\prime=\gamma_2-4b_0\, B_2\\
    \gamma_3& \to \gamma_3^\prime=\gamma_3-4b_1\, B_2
\end{aligned}
\end{equation}
\begin{equation}
\begin{aligned}
s_2\to&  s_2^\prime =s_2 +\frac{8b_0 s_0}{b_0+2\gamma_0} B_2 = -0.00040082+\frac{216}{175}B_2,\\
&\hspace{-1.cm}s_{2,i}\to s_{2,i}^\prime=s_{2,i} +\frac{216}{175} B_2 , \, i=\{1,3\}  \\
&(s_2^{nd})^\prime=s_2^{nd}  
\end{aligned}
\end{equation}
\begin{equation}
s_3\to s_3^\prime =s_3 +\frac{4B_2}{b_0+\gamma_0} \left(b_0 s_1 
-\frac{b_0}{4} (a_{10}+4a_{11}) +
2\frac{s_0}{b_0+2\gamma_0}(b_1 \gamma_0-b_0\gamma_1) \right)=-0.00008304-\frac{111}{280\pi^2}B_2
\end{equation}
where $a_{10} = -2(b_0-2\gamma_0) s_0$, $a_{11}=-(3/4) a_{10}$.

\section{Massive integrals and fitting functions}
\begin{figure}[h!]
    \hspace{-3cm}
    \begin{subfigure}[h]{0.33\textwidth}
        \epsfig{file=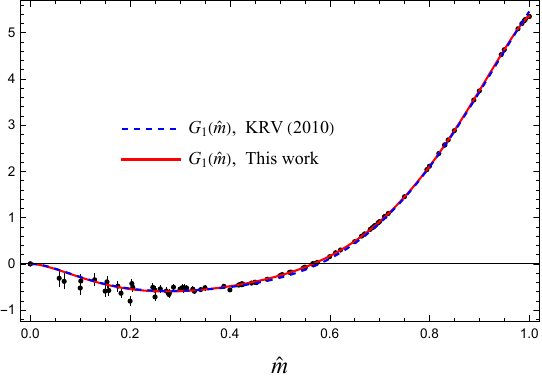,width=1.3\linewidth,angle=0}
        \label{Fig:G1comp}
    \end{subfigure}
    \hspace{1.5cm}
     \begin{subfigure}[h]{0.33\textwidth}
        \epsfig{file=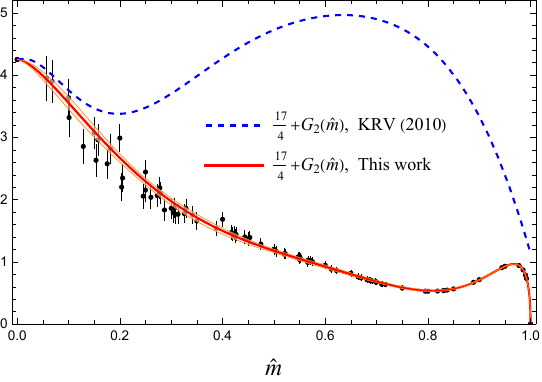,width=1.27\linewidth,angle=0}
        \label{Fig:G2comp}
    \end{subfigure}

    \hspace{-3.5cm}
     \begin{subfigure}[h]{0.33\textwidth}
        \epsfig{file=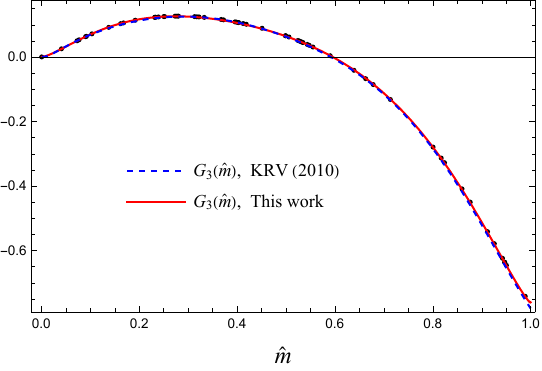,width=1.3\linewidth,angle=0}
        \label{Fig:G3comp}
    \end{subfigure}
    \hspace{1.5cm}
     \begin{subfigure}[h]{0.33\textwidth}
        \epsfig{file=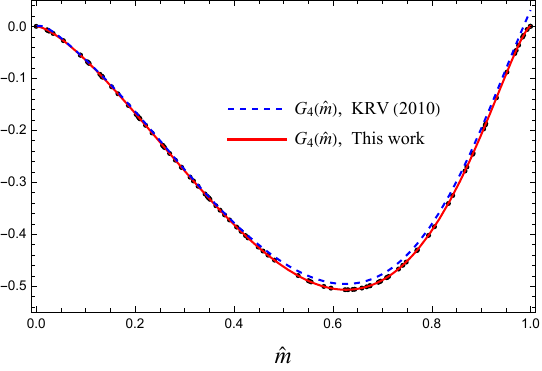,width=1.3\linewidth,angle=0}
        \label{Fig:G4comp}
    \end{subfigure}
    \caption{Comparison between the $G_i(\hat m=m/\mu)$ fitting functions from KRV (2010) \cite{Kurkela:2009gj} and our reevaluated ones. The extra constant for $G_2$ is the one appearing in parenthesis in \Eq{\ref{eq:PressureM3}} such that its sum with $G_2$ must vanish for $\hat{m}\to 1$. 
    The envelop shown for $G_2(\hat m)$ represents 3-$\sigma$ confidence level of the obtained fit, given statistical errors of ${\cal O}(\leq 1 \%)$ obtained for the relevant $I_k$ integral results. }
    \label{Fig:Gi_Comparison}
\end{figure}

Most basic integrals for the quark mass dependence were originally defined and evaluated in \cite{Kurkela:2009gj} or given as supplemental material in \cite{WebSiteRomatchke}. 
We give here some details on our reevaluation of the fitting functions $G_i(x)$ entering
Eq.(\ref{eq:PressureM3}), obtained as combinations of the basic massive integrals 
$I_k, k=1\cdots 12$ entering 2-cut and 3-cut contributions, Eqs.(\ref{2cut}),(\ref{3cut}). As mentioned in Sec. II, the $m\to 0$ limit of Eq.(\ref{eq:PressureM3})
is very nontrivial due to apparent $\ln(m)$ divergences, 
and the fitting procedure can be guided by the necessary cancellations of such IR divergent 
terms together with the known\cite{pQCDmu4L} $m\to 0$ limit of Eq.(\ref{eq:PressureM3}), and from inferring the $m\to 0$ limit
of some of the $I_k$. 
There are first analytically integrable ones:  
\be \label{I1}
I_1(\hat m) = \int \frac{d^3p}{(2\pi)^3} \frac{\Theta(\mu-E_p)}{2E_p} \equiv \int_p = \frac{\mu^2}{8\pi^2}\,z,
\ee
\be
I_2(\hat m) =\int_p \int_q \frac{1}{(P-Q)^2}=  \frac{\mu^2}{64\pi^2}\,\frac{\hat u^4-z^2}{\hat m^2} \underset{ m\to 0}{=}\ 
-\frac{\mu^2}{64\pi^2}\,\left(1+2\ln\(\frac{\hat m}{2}\)+{\cal O}(\hat m^2)\right)
\ee
\be
 I_{1b}(\hat{m})=\fr{\partial}{\partial \widetilde{m}^2}\int_p=\fr{1}{8\pi^2}\fr{z-\hat{u}}{\hat{m}^2}
\underset{ m\to 0}{=}\ 
\frac{1}{8\pi^2}\,\ln\(\frac{\hat m}{2}\)+{\cal O}(\hat m^2)
 \ee
\ba
I_{2b}(\hat{m})&=&\fr{\partial}{\partial \widetilde{m}^2}\int_p\int_q \fr{1}{(P-Q)^2}
 =\fr{1}{128\pi^4}\Bigg\{\({\rm arctan}\(\fr{\hat{u}}{\hat m}\)\)^2-\fr{2\hat u}{\hat m}{\rm arctan}\(\fr{\hat u}{\hat m}\) -\(\ln\(\fr{1+\hat u}{\hat m}\)\)^2\Bigg\}\nn
 &&\underset{ m\to 0}{=}\ \frac{1}{128\pi^4}
 \left(-\frac{\pi}{\hat m} +2+\frac{\pi^2}{4} -\ln^2\(\frac{\hat m}{2}\)
 +{\cal O}(\hat m) \right)
 \ea
 \ba
 I_8(\hat{m})&=&\int_p\int_q 
\left(1-\fr{E_q}{E_p}\right) \fr{1}{\((P-Q)^2\)^2}
=-\frac{1}{128 \pi ^4}\left(\frac{\hat u}{\hat m}-{\rm arctan}\left(\frac{\hat u}{\hat m}\right)\right)^2\nn
&&\underset{ m\to 0}{=}\ \frac{1}{128\pi^4}
 \left(-\frac{1}{\hat m^2} +\frac{\pi}{\hat m} -1-\frac{\pi^2}{4}
 +{\cal O}(\hat m) \right)
\ea
where $ \hat{m},\hat u, z$ were defined after Eq.(\ref{1cut}), and we give straightforward $m\to 0$ limits.
Note that in all integrals $P, Q,...$ are Euclidean on-shell four-momenta, i.e. $P^2=Q^2=-m^2$, and the remaining integrals $\int_p, \int_q$ are 
three-dimensional with $p_F$ as cutoff, as explicit in Eq.(\ref{I1}). 
Moreover, $\partial_{\widetilde m^2}$ outside the integral
acts only on the mass in the $p$-integration measure, after which $\widetilde m^2= m^2$ is set.
Remark also the useful relation:
\be
2(I_{2b}+I_8)-\fr{\partial\, I_2}{\partial m^2} =0 
\ee 
where now $\partial_{m^2}$ is the standard full derivative.
Next, the other relevant integrals can only be performed numerically, these are reproduced below
from \cite{Kurkela:2009gj} for convenience for the following discussion: 
 \ba
 I_{1c}(\hat{m})&=& \int_p\int_q \ln\(\frac{(P-Q)^2}{m^2}\)\\
 I_{2c}(\hat{m})&=& \int_p \int_q \frac{1}{(P-Q)^2 }\ln\(-\frac{(P-Q)^2}{m^2}\)\\
 I_3(\hat{m})&=&\int_p\int_q\int_r \fr{1}{(P-Q)^2(P-R)^2},\\
 I_{3b}(\hat{m})&=&\fr{\partial}{\partial \widetilde{m}^2}\int_p\int_q\int_r \fr{1}{(P-Q)^2(P-R)^2},\\
 I_4(\hat{m})&=&\int_p\int_q\int_r \fr{(Q-R)^2}{(P-Q)^2(P-R)^2},\\
 I_5(\hat{m})&=&\int_p\int_q\int_r \fr{1}{(P-Q-R)^2+m^2},\\
 I_6(\hat{m})&=&\int_p\int_q\int_r \fr{1}{(P-Q)^2(P-R)^2\left((P-Q-R)^2+m^2\right)},\\
 I_7(\hat{m})&=&\int_p\int_q\int_r \fr{(P-Q)^2}{(P-R)^2\left((P-Q-R)^2+m^2\right)},
 \ea
 \ba
 I_9(\hat{m})&=&\int_p\int_q\int_r \left(1-\fr{E_q}{E_p}\right) \fr{2}{\((P-Q)^2\)^2\, (P-R)^2} ,
 \ea
 where all integrals $I_3-I_7$ and $I_9$ require six-dimensional integration over three momenta $p,q,r$ and three angles. 
 The last three integrals entering the 2-cut contributions 
 are\footnote{There is a typo 
in Eq.(D16) of \cite{Kurkela:2009gj} for $I_{11}$, where the second term should be $32m^4I_2^\prime\to 32m^4I_{2b}$.}, defining $a=(P-Q)^2/m^2=-2+2(E_p E_q -p.q)/m^2$,
 \ba\label{I10}
 I_{10}(\hat{m})&=&\int_p\int_q\Bigg\{ -\frac{23}{3}+\frac{ 8 }{a}- \ln(a)+
 \sqrt{\frac{a}{4+a}}\Bigg[-\pi^2\(1+\frac{4}{3a^2}\) +\ln^2 (a) -2h_1^2(a)
  +\frac{8\ln(a)}{a^2} h_3(a)  \nn
 && +\(\frac{4}{a^2}-1\) \Big( \ln^2(4+a) -2h_2^2(a) +4h_5(a) \Big)
 +4h_4(a) \Bigg]
 +\sqrt{\frac{4+a}{a}}\left(\frac{4}{a}-3 \right)h_3(a)\Bigg\},
 \ea
 \ba\label{I11}
 I_{11}(\hat{m})&=&-16m^2 I_1 I_{1b}+32m^4 I_{2b} + \int_p\int_q\Bigg\{  12+\frac{32}{a^2}
 (1-\frac{E_q}{E_p})
 +\left(2-\frac{8}{\pmq}\right) \ln(a) +\sqrt{\frac{a}{4+a}}\Bigg[\left( 2+\frac{16}{a}+\frac{8}{a^2}\right)\frac{\pi^2}{3}\nn &&-2 \ln^2(a) +4h_1^2(a)-\frac{16\ln(a)}{a^2} h_3(a) +\left(\frac{4}{a^2}-1\right)
 \Big(4 h_2^2(a) -2\ln^2\left(4+a\right) -8h_5(a)\Big) \nn
 && +\left(\frac{4}{a}-1\right) \Big( 4h_6(a) +h_3^2(a)\Big) -8h_4(a)\Bigg]+\sqrt{\frac{4+a}{a}}\left(6-\frac{8}{\pmq}\right)
 h_3(a)
 \Bigg\},
 \ea
 \ba\label{I12}
 I_{12}(\hat{m})&=&\int_p\int_q \left(\frac{a-2}{a^2}\right)\(4-a\ln(a)+2(a-2)\sqrt{\frac{4+a}{a}}{\rm Arctanh}\[\sqrt{\frac{a}{4+a}}\]\),
\ea
where we introduced 
 \ba
&& h_1(a)=\ln\left(\frac{1}{2}\left[\sqrt{a(4+a)}-a \right]\) ,\nn
&& h_2(a)=\ln\(\frac{1}{2}\left[4+a-\sqrt{a(4+a)}\right]\) , \nn
&& h_3(a)=\ln \left( \frac{1}{2}\left[2+a-\sqrt{a(4+a)}\right]\right) , \nn
&& h_4(a)= \text{Li}_2\(\frac{1}{2} \left[1-\sqrt{\frac{4+a}{a}}\right]\) ,\nn
&& h_5(a)= \text{Li}_2\(\frac{1}{2}\left[1-\sqrt{\frac{a}{4+a}}\right]\) , \nn
&& h_6(a)= \text{Li}_2\( \frac{1}{2}\left[2+a-\sqrt{a(4+a)}\right]\).
\ea
Eqs.(\ref{I10})-(\ref{I12}) are to be integrated over $p, q$ and $z\equiv p.q/(p q)$.
It is not difficult to extract the divergent pieces for $m\to 0$ of
(\ref{I10}),(\ref{I11}) as
\ba
&& I_{10}(m) \underset{ m\to 0}{\simeq} -(\frac{23}{3}
+\pi^2)I_1^2(m\to 0) +2I_{1c}(m\to 0) +I_{10}^{f}(m)\nn
&& I_{1c}(m) = I_{1c}^f(m)-2 \ln\(\frac{\hat m}{2}\)I_1^2(m)
\ea
\vspace{-0.2cm}
\ba
&&I_{11}(m) \underset{ m\to 0}{\simeq} (12
+\frac{2\pi^2}{3})I_1^2(m\to 0) -4I_{1c}(m\to 0) -I_{1d}(m\to 0) +I_{11}^{f}(m)\nn
&&I_{1d}(m) =\int_p\int_q \ln^2\(\frac{(P-Q)^2}{m^2}\) = I_{1d}^f(m)-4 \ln\(\frac{\hat m}{2}\) I_{1c}^f(m)+
4 \ln^2 \(\frac{\hat m}{2}\) I_1^2(m)
\ea
where the $I_k^f$ are finite integrals for $m\to 0$.\\
Accordingly, one can infer that the following combinations 
of basic integrals $I_k$ should be finite for $m\to 0$:
\be \label{fincomb}
\begin{split}
\mathcal{M}_3^{2c}:&\ C_A\(\frac{5}{3} I_{1c}-\frac{22}{3}I_1^2\Ll{}+I_{10}\)\\
\mathcal{M}_3^{2c}:&\ N_f\(\frac{4}{3}I_1^2\Ll{}-\frac{2}{3}I_{1c}\)\\
\mathcal{M}_{3}^{3c}:&\ (C_A+C_F)\(2I_1 I_2-4(I_5+I_7) \)\\
\mathcal{M}_{3}^{2c}\ {\rm and}\ \mathcal{M}_{3}^{3c}:&\  C_F\((4\pi)^2 I_{11}+2I_{1}^2I_{1b}-2I_4\). 
\end{split}
\ee
We have reevaluated the 
basic $I_k$ with accurate multidimensional Monte-Carlo integration 
methods\cite{vegas,Cuba}, for which we find excellent agreement with \cite{Kurkela:2009gj}\footnote{Reaching about $1\%$ relative accuracy at worst for individual $I_k$ integrals, 
which is comparable to results given in \cite{WebSiteRomatchke} used in \cite{Kurkela:2009gj}. Our resulting fitting functions for all individual $I_{k}$ integrals can be provided
upon request.}.
From these we obtain the new determination of the fitting functions 
$G_i,\ i\in\{1,2,3,4\}$ given in Eq.(\ref{eq:GiFunc}), having used available data in \cite{WebSiteRomatchke} and extra data that we determined to match a bit more precisely the particularly nontrivial $m\to 0$ limit,
as well as the $m\to\mu$ limit.
The $G_i(\hat m)$ of Eq.(\ref{eq:GiFunc}) are numerically 
illustrated in Fig. \ref{Fig:Gi_Comparison}, also
compared to the similar functions
in \cite{Kurkela:2009gj}. 
The error bars for $G_3$ and $G_4$ are too small to be seen in the figure
scale, and the propagation of uncertainties from $G_2$ to the pressure is rather negligible.\\

According to Fig. \ref{Fig:Gi_Comparison}, we thus obtain very good agreement for $G_1$, $G_3$, $G_4$, but sizable differences in $G_2(\hat m)$  as compared to \cite{Kurkela:2009gj} for intermediate 
and large $\hat m=m/\mu $ values\footnote{These $G_2(x)$ discrepancies appear 
to be mainly due to a typo in a numerical 
code (A. Kurkela, A. Vuorinen, personal communication).}.
Notice also the vanishing of the $C_F$ contribution $17/4+G_2(\mh)\to 0$ for  $\mh\to 1$ in Fig. \ref{Fig:Gi_Comparison},
as expected for $\mathcal{M}_3$ consistently from the fact that $p_F$ provides a cut-off for the momenta integration domain.
\section{non-diagonal massive NNLO contributions}
\subsection{Derivation of the NNLO non-diagonal $\mathcal{M}_3^{N_f=2^*+1^*}$ contribution}
For the sake of clarity we stick to the same notations introduced in \cite{Kurkela:2009gj}.
The one-loop gluon polarization tensor is conveniently split into a vacuum ($T=\mu=0$) and matter contribution via
\ba
\Pi^{\mu\nu}(\vec{m}^2,K)&=&\Pi_{\rm vac}^{\mu\nu}(\vec{m}^2,K)+\Pi_{\rm mat}^{\mu\nu}(\vec{m}^2,K),
\ea
where we suppressed the trivial color factor $\delta_{ab}$ and $\vec{m}^2$ means $(m_i^2)_{i\in 1\dots N_f}$. Lorentz symmetry and gauge invariance give 
\ba
\Pi^{\mu\nu}_{\rm vac}(\vec{m}^2,K)&=&\Pi_{\rm vac}\(\vec{m}^2,K^2\)\fr{g^2}{(4\pi)^2}\(\fr{\Lambda^2}{K^2}\)^{\!\!\e}\(K_{\mu}K_{\nu}-
K^2\delta_{\mu\nu}\), \label{polarvac}
\ea
and a direct calculation leads to
\ba
\Pi_{\rm vac}\(\vec{m}^2,K^2\)&=&\underbrace{-2^{5-2d}\pi^{\frac{7-d}{2}}(3d-2)C_A\fr{{\rm csc}(\frac{\pi\,d}{2})}{\Gamma(\frac{d+1}{2})}}_{f_0}-\sum_{j=1}^{N_f}2^{6-d}\pi^{\frac{4-d}{2}}\Gamma\(\frac{4-d}{2}\)\int_0^1{\rm d}x (x(1-x))^{\frac{d-2}{2}}\(1+\fr{m_j^2/K^2}{x(1-x)}\)^{\frac{d-4}{2}}\nn
&\equiv& f_0+\sum_{j=1}^{N_f}f_1\(\fr{m_j^2}{K^2}\).\label{f0f1}
\ea
To evaluate $\mathcal{M}_3^{N_f=2^*+1^*}$, we only need to modify the VM graph which only contributes to the 1-cut and 2-cut contributions. The expressions for the 2GI graphs derived in \cite{Kurkela:2009gj} are 
\ba
\frac{\Omega_{\rm 2GI}^{m,1c}(m_i)}{V}&=&-d_A m_i^2\Bigg\{C_A\Bigg[\frac{16}{\e^2}+\frac{10}{3}\(6\ln\frac{\Lambda}{m_i}+1\)\frac{1}{\e}\nn
&-&\(4\ln\frac{\Lambda}{m_i}+\frac{136}{3}\)\ln\frac{\Lambda}{m_i}-\frac{82}{3}+\pi^2\(\frac{41}{6}-8\ln\,2\)+12\zeta(3)\Bigg]I_1(m_i)\nn
&+&C_F\Bigg[\bigg\{\frac{18}{\e^2}-\frac{3}{2\e}
-6\(12\ln\frac{\Lambda}{m_i}+5\)\ln\frac{\Lambda}{m_i}-\frac{313}{4}\nn
&-&\pi^2\(\fr{35}{3}-16\ln\,2\)-24\zeta(3)\bigg\}I_1(m_i)\nn
&+&m_i^2 \bigg\{\frac{18}{\e^2}-36\(2\ln\fr{\Lambda}{m_i}+\frac{8}{3}\)\ln\frac{\Lambda}{m_i}-32\bigg\}I_{1b}(m_i)\Bigg] \label{lnZ2GI1c}\\
&-&N_f\Bigg[\frac{4}{\e^2}+8\bigg(\ln\frac{\Lambda}{m_i}+\frac{2}{3}\bigg)\frac{1}{\e}
+8\(\ln\frac{\Lambda}{m_i}+\fr{4}{3}\)\ln\frac{\Lambda}{m_i}+\fr{32}{3}+\frac{\pi^2}{3}\Bigg]I_1(m_i)\Bigg\}\,\frac{g^4({\Lambda})}{(4\pi)^4},\nn
\fr{\Omega_{\rm 2GI}^{m,2c}(m_i)}{V}&=&-d_A \Bigg\{ C_A\left[
\left(\frac{5}{3}I_1(m_i)^2-\frac{10}{3}m_i^2 I_2(m_i)\right)/\e + I_{10}(m_i)
+ (-4I_1(m_i)^2+8 m_i^2 I_2(m_i))\ln\fr{\bar\Lambda}{m_i}\right]\nonumber\nn
&+&C_F\left[I_{11}(m_i)+\left[24(m_i^2 I_2(m_i) -m_i^2 I_{1b}(m_i)I_1(m_i)+2m_i^4 I_{2b}(m_i))+48 m_i^4I_8(m_i)\right]\ln\fr{\Lambda}{m_i}\right]\nn
&+&N_f \Bigg[\left(-\frac{2}{3}I_1(m_i)^2+\frac{4}{3}m_i^2 I_2(m_i)\right)/\e + \frac{2}{3}I_1(m_i)^2\Bigg]\Bigg\}\,\fr{g^4({\Lambda})}{(4\pi)^2}, \label{lnZ2GI2c}\\
\frac{\Omega_{\rm 2GI}^{m,3c}(m_i)}{V} &=& -{\cal M}_3^{3c}(m\to m_i)
\,\frac{g^4(\Lambda)}{(4\pi)^4} \label{lnZ2GI3c}
\ea
where ${\cal M}_3^{3c}$ is given in Eq.(\ref{3cut}) and the coupling $g$ and the quark mass $m_i$ are renormalized quantities. The divergences in the previous contributions cancel out with lowest order mass and coupling
renormalization (see the discussion in Appendix A). There is no need to modify the $N_f$ coefficient depending on the case at hand ($N_f=2+1^*,\ N_f=3^*,\dots$) in the 2GI function since it only comes from renormalization.
\subsection{Non-diagonal VV contribution}\label{AppendixC2}
The ``VV'' additional contribution to the pressure, for arbitrary  $m_i, m_j$ different quark masses,
may be derived from appropriately applying RG invariance properties with already known
three-loop vacuum results from specific quantities, thus without need of actual three-loop
calculations. More precisely, using results for the (quark sector) vacuum energy anomalous
dimension\cite{vacEn5l}, and from the quark 
condensate\cite{KNcond,KNcond2},
we can obtain the non-diagonal NNLO VV pressure contribution. The vacuum pressure given in \Eq{\ref{eq:PNNLOrgopt}} is 
normalized to one flavor of quark, but here it is more convenient to rewrite this contribution directly taking into account the $N_f \otimes N_f $ degrees of freedom. Using combinatorics, one finds:
\begin{equation}\label{eq:PvacNNLOmix}
\begin{aligned}
 N_f\,P^{v}_{\rm NNLO}(m,N_h=N_f,N_l=0)\to  P^{v}_{\rm NNLO}\equiv & (N_f-1)P^{v,d}_{\rm NNLO}(m,N_h=N_f-1,N_l=1)\\
 &+P^{v,d}_{\rm NNLO}(m_3,N_h=1,N_l=N_f-1)+2(N_f-1)P^{v,nd}_{\rm NNLO}(m,m_3),\\
 \end{aligned}
 \end{equation}
 with 
 \begin{equation}
 \begin{aligned}
     P^{v,nd}_{\rm NNLO}(m,m_3)=& \frac{1}{\pi^2} \frac{g^4}{(4\pi)^4} 
 m^2 m_3^2 
 \left( \alpha_{32}^{nd} (\ln \frac{m}{\Lambda} +\ln \frac{m_3}{\Lambda} )+
 \alpha_{33}^{nd} \right)\\
 \alpha_{32}^{nd} =& -3 N_c C_F T_f =-6, \;\;\;
 \alpha_{33}^{nd} = \frac{29}{3} -12\zeta(3) . 
 \end{aligned}
 \end{equation}
 In the diagonal part of the pressure, $N_l$ has been replaced by $N_l=N_f-N_h$ such that $N_f=N_h+N_l$ is always satisfied. As a crosscheck, for $\mcm =0$, i.e, $m_3=m$, the right hand-side of \Eq{\ref{eq:PvacNNLOmix}} reproduces the left-hand side. 
\subsection{Non-diagonal VM contribution}
Following \cite{Kurkela:2009gj}, we re-evaluate the 1- and 2-cut contributions: 
\begin{equation}\label{eq:VM_1c_step1}
\begin{split}
    \frac{\Omega_{\rm VM}^{\rm m,1c}(m_i)}{V}=&\ 2d_A \frac{g^4}{(4\pi)^2}\Lambda^{2\e}\Bigg\{ 2m_i^2\left(f_0 \DqC{m_i}{1}+\sum_{j}^{N_f}\DqTC{m_j}{m_i}{1}\right)\\
    &-(1-\e)\left(f_0\DqC{m_i}{\e}+\sum_{j}^{N_f}\DqTC{m_j}{m_i}{\e}\right)+(1-\e)\sum_{j}^{N_f}\DcT{m_j}{0}{1+\e}\Bigg\},
\end{split}    
\end{equation}
\begin{equation}\label{eq:VM_2c_step1}
\begin{split}
    \frac{\Omega_{\rm VM}^{\rm m,2c}(m_i)}{V}=& -d_A \frac{g^4}{(4\pi)^2}\Lambda^{2\varepsilon}\Bigg\{ 2m_i^2\left(\frac{f_0}{((P-Q)^2)^{1+\e}}+\frac{1}{(P-Q)^2}\sum_{j}^{N_f}\DqTCC{m_j}{m_i}\right)\hspace{2.2cm}\\
    &-(1-\e)\left(\frac{f_0}{((P-Q)^2)^{\e}}+\sum_{j}^{N_f}\DqTCC{m_j}{m_i}\right)\Bigg\},
\end{split}    
\end{equation}
where $m_j$ is the mass of the quark flowing in the vacuum loop and $m_i$ the mass flowing in the matter loop. The tilde on the $D_i$ functions means that it is multiplied by $f_1(m_j)$ prior to integration.
For this calculation we needed to reevaluate the 1-cut $\mathcal{D}_{4}$ integrals whose expression are given in Eqs.(\ref{eqC14}),(\ref{eqC16}). \\
Once implemented in $\Omega_{\rm VM}$ we find:
\begin{equation}\label{eq:VM1cPlusFnui}
    \begin{split}
       \hspace{-1.5cm} \frac{\Omega_{\rm VM}^{m,1c}(m_i)}{V}=&\frac{d_A m_i^2 g^4 I_1(m_i)}{(4\pi)^4}\Bigg\{C_A\Bigg[\frac{5}{\e^2}+\frac{1}{\e}\left(20\Ll{i}+\frac{39}{2}\right)+2\left(20\Ll{i}+39\right)\Ll{i}+\frac{261}{4}+\frac{25\pi^2}{6}\Bigg]-N_f\Bigg[\frac{2}{\e^2} +\frac{1}{\e}\left(7+8\Ll{i}\right)\\
       &+\frac{13}{2}+\frac{\pi^2}{3}+16\Llc{i}+28\Ll{i}\Bigg]-\sum_{j=1}^{N_f}\Bigg[4\left({\rm Li}_{2}(\nui)\(1+\nui^{-2}\)-\frac{1}{\nui^2}-\Phi(\nui,2,\frac{3}{2})(1+\nui)+\mathcal{F}(\nui)\right)\Bigg]\Bigg\},\\
        &\mathcal{F}(\nui)=\ln\nui\Bigg\{\ln(1-\nui)\left(1+\nui^{-2}\right)+2\frac{{\rm Arctanh}(\sqrt{\nui})}{\sqrt{\nui}}(1+\nui^{-1})-\nui^{-1}-\frac{1}{2}\ln\nui\Bigg\},\ \ \mathcal{F}(\nui=1)=0,
    \end{split}
\end{equation}
with  $\nui=(m_i/m_j)^2$, and
\begin{equation}\label{OmVM2cbare}
    \begin{split}
      \hspace{-1cm} \frac{\Omega_{\rm VM}^{m,2c}(m_i)}{V}=& \frac{d_A g^4}{(4\pi)^2}\Bigg\{C_A\Bigg[\frac{1}{\e}\left(\frac{5}{3}I_1(m_i)^2-\frac{10}{3}m_i^2 I_2(m_i)\right)+\left(\frac{10}{3}I_1(m_i)^2-\frac{20}{3}m_i^2 I_2(m_i)\right)\Ll{i}+\frac{16}{9}I_1(m_i)^2-\frac{62}{9}m_i^2 I_2(m_i)\\
     & -\frac{5}{3}I_{1c}(m_i)+\frac{10}{3}m_i^2 I_{2c}(m_i)\Bigg]-N_f\Bigg[\frac{1}{\e}\left(\frac{2}{3}I_1(m_i)^2-\frac{4}{3}m_i^2 I_2(m_i)\right)+\left(\frac{4}{3}I_1(m_i)^2-\frac{8}{3}m_i^2 I_2(m_i)\right)\Ll{i}\\
     &+\frac{4}{9}I_1(m_i)^2-\frac{20}{9}m_i^2 I_2(m_i)-\frac{2}{3}I_{1c}(m_i)+\frac{4}{3}m_i^2 I_{2c}(m_i)\Bigg]+\frac{2}{3} \sum_{j=1}^{N_f}I_{12}^{\rm bis}(m_i,\nui)\Bigg\}.
    \end{split}
\end{equation}
\subsection{Unequal mass integrals}
Finally additional ingredients needed for our generalization 
to unequal quark masses, entering Eq.(\ref{OmVM2cbare}), are 
\be
\label{eqC13}
\begin{aligned}
\hspace{-1cm}I_{12}^{\rm bis}(\hat{m}_i,\nui)=&\int_p\int_q \left(\frac{a_i-2}{a_i^2\,\nui}\right)\(4-a_i\,\nui\ln(a_i\,\nui)+2(a_i\,\nui-2)\sqrt{\frac{4+a_i\,\nui}{a_i\,\nui}}{\rm Arctanh}\[\sqrt{\frac{a_i\,\nui}{4+a_i\,\nui}}\]\),\ \ a_i=\frac{\(P-Q\)^2}{m_i^2}\\
=\  \mh^2 \uh^2  \Bigg\{&-3.459417 + 0.421797  \mh^2 - 
   0.776318 \uh-0.770475\uh^3 +(0.387893 + 2.123092\mh)\nui\\
   &+ (0.181022 - 0.506166\mh +0.139366\mh^2)\nui^2 +(0.664467 + 0.115689\mh + 0.115689\ln\mh)/(\mh\ \nui)\\
   &+ (1.960279 - 3.332848  \mh) \ln\nui + (-0.585635 - 
      0.680364  \mh + 0.576210  \ln\mh)  \ln^2\nui\Bigg\}.
\end{aligned}
\ee
 Note that the 2-dimensional fit of $I_{12}^{\rm bis}(\mh_i,\nui)$ is only valid (within $\sim 1\%$ accuracy) in the region $x \otimes \nui \in [0,1]\otimes [0.4,2.5]$,
 where the latter range is largely sufficient for physically relevant values of $\nui$. \\
 The last ingredients required to generalize the ``VM'' contribution for the
 $N_f=2^*+1^*$ case are the vacuum amplitude $\Tilde{D}_i$, defined in Fig. 12 in \cite{Kurkela:2009gj} and obtained from \cite{Di_Ref}. These functions correspond to different scalar Feynman graph topologies, where an upper index indicate the mass of the propagator and the lower index indicates the power of the propagator. For tilde functions, the first dual arguments (before the vertical line) indicates the propagator coming from the multiplication of $f_1$ prior to integration :
 \be\label{eqC14}
  \begin{aligned}
   \mathcal{D}_{4}=& \ \begin{tikzpicture}[baseline=-\the\dimexpr\fontdimen22\textfont2\relax]
            \begin{feynman}
                 \draw [domain=0:360] plot ({0.8*cos(\x)}, {0.8*sin(\x)});
                 \vertex (1) at ($(0,0.9)$){};
                 \vertex (2) at ($(0,-0.9)$){};
                 \draw (1)--(2);
            \end{feynman}
            \end{tikzpicture}\ \ , \ \ \ \mathcal{D}_{5}=\begin{tikzpicture}[baseline=-\the\dimexpr\fontdimen22\textfont2\relax]
            \begin{feynman}
                 \draw [domain=0:360] plot ({0.7*cos(\x)}, {0.7*sin(\x)});
            \end{feynman}
            \end{tikzpicture}\\
\DqCfull{m_1}{m_2}{0}{\alpha}{\beta}{\gamma}=\int_{-\infty+i\mu}^{\infty+i\mu}dp_0\int\frac{d^{d-1}p}{\(2\pi\)^{d-1}}&\int_{-\infty+i\mu}^{\infty+i\mu}dq_0 \int\frac{d^{d-1}q}{\(2\pi\)^{d-1}}\frac{1}{\(p_0^2+p^2+m_1^2\)^\alpha \(q_0^2+q^2+m_2^2\)^\beta \((p_0-q_0)^2+(p-q)^2\)^\gamma}.\\
\end{aligned}
 \ee
 For the relevant $D_i$ function we use dimensional regularization where the integration measure reads:
\begin{equation}
    \int_{P}=\left(\frac{\Lambda^2 e^{\gamma}}{4\pi}\right)^{\e}\int\frac{d^d P}{(2\pi)^d},\ \ \ \int_{\Vec{p}}=\left(\frac{\Lambda^2 e^{\gamma}}{4\pi}\right)^{\e}\int\frac{d^{d-1} \Vec{p}}{(2\pi)^{d-1}}, \ \ \ d=4-2\e,
\end{equation}
where $P$ is an Euclidean momentum and $\gamma$ is Euler's constant. 
 On top of that, a ``$c$'' lower index indicates that the corresponding propagator has been cut following the rules explained in section II.A.2.\\
 Finally, the $\tilde{\mathcal{D}_i}$ which required re-evaluation when 
 using two different masses, reads
\begin{equation}\label{eqC16}
    \begin{split}
         &\left(\frac{\Lambda^2 e^{\gamma}}{4\pi}\right)^{\e}\DqTC{m_j}{m_i}{1+\e}=-\frac{1}{3(4\pi)^2}\left(\frac{\Lambda^2}{m_i\ m_j}\right)^{2\e}\Bigg\{\frac{1}{\e^2}+\frac{1}{\e}\left(\frac{14}{3}-\ln\nui\right)+(1-3\nui)\Phi(\nui,2,\frac{3}{2})+\frac{46}{9}+\frac{\pi^2}{6}\\
        &\hspace{4cm}+\frac{2}{3}\ln\nui\Bigg(-7+3\nui^{-1}+\nui^{-\frac{1}{2}}(9-3\nui^{-1}){\rm Arctanh}(\sqrt{\nui})+3\ln(1-\nui)\Bigg)+2{\rm Li}_{2}(\nui)\Bigg\}\\
         &\left(\frac{\Lambda^2 e^{\gamma}}{4\pi}\right)^{\e}\DqTC{m_j}{m_i}{\e}=\frac{m_i^2}{3(4\pi)^2}\left(\frac{\Lambda^2}{m_i\ m_j}\right)^{2\e}\Bigg\{\frac{1}{\e^2}+\frac{1}{\e}\left(\frac{13}{6}+6\nui^{-\frac{1}{2}}-\ln\nui\right)-\frac{67}{36}+\frac{\pi^2}{6}-8\,\nui\Phi(\nui,2,\frac{5}{2})\\
        &\hspace{2cm}+\frac{1}{2}\ln(\nui)\left(-\frac{13}{3}-\frac{8}{\nui}-\ln(\nui)+\frac{32}{(\sqrt{\nui})^3}{\rm Arctanh}(\sqrt{\nui})+\left(4+\frac{12}{\nui^2}\right)\ln(1-\nui)\right)+\left(2+\frac{6}{\nui^2}\right){\rm Li}_{2}(\nui)\Bigg\}.
    \end{split}
\end{equation}
For this calculation we used Mellin-Barnes techniques to perform the integration over the Feynman parameters (see for instance \cite{Smirnov:2012gma} for a review). In practice, it amounts to factorize the mass dependence of the propagator coming from $f_1$ in $\tilde{\mathcal{D}}_4$ into:
\be
\(1+\frac{m^2}{(P-Q)^2x(1-x)}\)^{-\e}=\frac{1}{\Gamma(\epsilon)} \,\frac{1}{2\pi i}\oint ds\, \Gamma(-s)\Gamma(s+\e)\(\frac{m^2}{\(P-Q\)^2\, x(1-x) }\)^s,
\ee
where $x$ is a Feynman parameter. Such factorization makes the integration over all Feynman parameters straightforward in terms of $\Gamma$ functions, then the remaining $s$-integral is carried out via the residue theorem with appropriately chosen contours.

\end{document}